\newif\ifcopernicus
\copernicusfalse

\ifcopernicus
    \documentclass[ascmo, manuscript]{copernicus}
\else
    \documentclass[11pt]{article}
    \usepackage[margin=1in]{geometry}
    \usepackage{authblk}
    \usepackage{setspace}      
\fi

\usepackage{natbib,hyperref}
\usepackage{graphicx,bm,colonequals,amsmath,amssymb,url,xcolor,bbm}
\usepackage{array,tabularx,multirow}
\usepackage[font={footnotesize}]{caption,subcaption}
\usepackage[utf8]{inputenc}
\usepackage{enumitem}
\usepackage{soul} 
\usepackage{placeins} 
\usepackage[normalem]{ulem}
\usepackage{tabularray} 
\UseTblrLibrary{amsmath} 
\usepackage{comment} 
\usepackage[ruled,vlined]{algorithm2e}
\usepackage{rotating}
\usepackage{tikz}
\usepackage{caption}
\usepackage{subcaption} 
\usepackage{tikz}
\usepackage{lineno} 
\usetikzlibrary{positioning,calc} 
\graphicspath{{plots/}}

\usepackage{mathtools}
\mathtoolsset{showonlyrefs} 
\everydisplay{\textstyle}
\bibpunct[, ]{(}{)}{;}{a}{,}{,}
\usepackage{amsthm}
\newtheoremstyle{propstyle} 
    {2mm}                    
    {1mm}                    
    {\itshape}                   
    {}                           
    {\scshape}                   
    {.}                          
    {.5em}                       
    {}  

\theoremstyle{propstyle}

\theoremstyle{propstyle}

\theoremstyle{propstyle}

\theoremstyle{propstyle}

\theoremstyle{propstyle}

\usepackage[ruled,vlined]{algorithm2e}
\SetKwInput{KwInput}{Input}
\usepackage{algorithmic}
\usepackage{array,framed}
\usepackage{float}

\newcommand{\bd}{\mathbf{d}}
\renewcommand{\bf}{\mathbf{f}}
\newcommand{\bs}{\mathbf{s}}

\newcommand{\bx}{\mathbf{x}}
\newcommand{\by}{\mathbf{y}}

\newcommand{\br}{\mathbf{r}}

\newcommand{\bT}{\mathbf{T}}

\newcommand{\bY}{\mathbf{Y}}

\newcommand{\bK}{\mathbf{K}}
\newcommand{\bX}{\mathbf{X}}

\newcommand{\bt}{\mathbf{t}}

\newcommand{\bftheta}{\bm{\theta}}

\newcommand{\tr}{tr}
\newcommand{\GP}{\mathcal{GP}}

\newcommand{\normal}{\mathcal{N}}

\DeclareMathOperator*{\argmax}{arg\,max}

\newcommand{\rev}[1]{\textcolor{black}{#1}}

\ifcopernicus

\title{Scalable \rev{stochastic} modeling of non-Gaussian spatio-temporal fields via autoregressive Gaussian processes}

\Author[1]{Carrie J}{Lei-Cramer} 
\Author[2]{Jian}{Cao}
\Author[3][katzfuss@gmail.com]{Matthias}{Katzfuss}

\affil[1]{Department of Statistics, Texas A\&M University, College Station, USA}
\affil[2]{Department of Mathematics, University of Houston, Houston, USA}
\affil[3]{Department of Statistics, University of Wisconsin--Madison, Madison, USA}

\runningtitle{Scalable stochastic modeling of non-Gaussian spatio-temporal fields}
\runningauthor{Lei-Cramer et al.}

\received{}
\pubdiscuss{}
\revised{}
\accepted{}
\published{}
\firstpage{1}

\else  

\title{Scalable \rev{stochastic} modeling of non-Gaussian spatio-temporal fields via autoregressive Gaussian processes}

\author[1]{Carrie J. Lei-Cramer}
\author[2]{Jian Cao}
\author[3]{Matthias Katzfuss}

\affil[1]{Department of Statistics, Texas A\&M University, College Station, USA}
\affil[2]{Department of Mathematics, University of Houston, Houston, USA}
\affil[3]{Department of Statistics, University of Wisconsin--Madison, Madison, USA Corresponding author: \href{mailto:katzfuss@gmail.com}{katzfuss@gmail.com}}

\date{}  

\fi

\begin{document}

\maketitle

\begin{abstract}
\rev{Modeling the full joint distribution of spatio-temporal fields} is crucial for a variety of applications, including stochastic weather generators and climate-model surrogates. However, many such fields exhibit complex dependence structures that vary across space and time and are nonlinear, resulting in nonstationary and non-Gaussian joint distributions. Our approach represents the joint density of a spatio-temporal field as a product of univariate conditional distributions and models these conditionals using Gaussian processes within an autoregressive transport-map construction. This prior distribution provides regularization, making our method suitable for a small number of training samples. Data-dependent sparsity in the conditioning sets ensures scalability to high-dimensional distributions. We also propose a variant of the method designed to sample or predict forward in time from a given incomplete space-time trajectory. We demonstrate the accuracy and scalability of our approach on non-Gaussian climate-model output with tens of millions of data points.
\end{abstract}

\section{Introduction}  

Spatio-temporal fields play a significant role in many scientific applications, such as climate modeling and environmental studies, where high-dimensional variables are observed across space and time. To highlight the need for effective statistical models, consider measurements of global surface temperatures collected across millions of spatial locations over a sequence of time points. Generating conditional samples or forecasting future states on such high-dimensional fields without knowing their true joint distribution is highly challenging. Furthermore, training data in the form of independent ensembles is often scarce. 
\rev{Running high-fidelity, physics-based climate models can require thousands of core-hours on high-performance computing systems to generate a single ensemble member.}
Therefore, it is essential to develop statistically principled, \rev{stochastic} surrogates that can accurately represent the joint distribution of spatio-temporal data to ensure dependable predictions, conditional forecasts, and uncertainty quantification, all while relying on a small amount of training data. Accurately inferring this joint distribution is particularly challenging when faced with intricate, nonstationary dependencies and non-Gaussian attributes. 

Many traditional methods for spatio-temporal analysis were originally designed for inference based on a single training sample and assume Gaussian processes (GPs) with straightforward parametric covariance functions \citep[e.g.,][]{Cressie1993}. While extensions to nonparametric covariances have been proposed \citep[e.g.,][]{Huang2011, Choi2013}, these approaches typically still rely on assumptions of Gaussianity. On the other hand, generative machine-learning approaches, such as generative adversarial networks or variational autoencoders \citep{Goodfellow2016, Kovachki2020}, can capture complex non-Gaussian distributions. However, they typically require a massive number of training samples, lack formal uncertainty quantification, and can be highly sensitive to the choice of tuning parameters and network architectures \citep{Arjovsky2017, Hestness2017}. \rev{The method we develop here also specifies and samples from a full joint density over the field. It differs from these deep-learning approaches in that inference is available in closed form, predictive uncertainty is quantified, and the model is fit from tens of ensemble members rather than from the large training corpora such networks require.}

To address the limitations of Gaussian assumptions without requiring massive training datasets, triangular transport maps \citep{Marzouk2016} offer a powerful framework. These maps autoregressively transform a complex, non-Gaussian target distribution into a simple reference distribution, typically a standard Gaussian. Building on this concept, \citet{Katzfuss02042024} introduced a Bayesian nonparametric approach for purely spatial fields, known as spatial autoregressive Gaussian processes (ARGPs). In this approach, the components of the autoregressive mapping are modeled using GPs. This allows for closed-form inference that effectively quantifies uncertainty and prevents overfitting even with limited training samples. To handle high-dimensional spatial data, the ARGP leverages a screening effect through nearest-neighbor conditional independence assumptions, effectively providing a non-Gaussian, nonparametric extension of Vecchia approximations \citep{Vecchia1988, Katzfuss02042024}. \citet{chen2024precision} showed that such sparsity assumptions allow for accurate estimation of a target distribution based on a polylogarithmic number of samples in the Gaussian setting.

In this paper, we propose a scalable, \rev{stochastic} modeling framework for non-Gaussian spatio-temporal fields. Our primary contributions are threefold:
\begin{enumerate}
    \item We generalize the spatial autoregressive Gaussian process (ARGP) transport-map framework of \citet{Katzfuss02042024} to spatio-temporal fields. By introducing \rev{data-driven} scaling factors for space-time coordinates, our method automatically balances spatial and temporal dependencies to determine appropriate conditioning sets.
    \item \rev{Within this scaled geometry, we propose two ordering strategies: a global space-time maximin ordering for unconditional simulation, and a chronology-respecting time ordering for probabilistic forecasting.}
    \item We show that the resulting method remains computationally feasible in extremely high-dimensional settings through data-dependent sparsity, successfully modeling a non-Gaussian spatio-temporal field in over 1.6 million dimensions based on a small ensemble of training samples.
\end{enumerate}

The remainder of the paper is organized as follows. In Section~\ref{sec:review}, we provide a brief review of (purely) spatial autoregressive Gaussian processes. In Section \ref{sec:methodology}, we describe our novel methodology for spatio-temporal fields. In Section \ref{sec:application}, we apply the methods to climate-model output, including demonstrating the scalability on a massive global dataset. Finally, we conclude in Section \ref{sec:discussion} with a summary of our findings and a discussion of future extensions.

\section{Review of spatial autoregressive Gaussian processes\label{sec:review}}

In this section, we briefly review the spatial autoregressive Gaussian process (ARGP) framework introduced by \citet{Katzfuss02042024}. We refer readers to their work for more comprehensive mathematical details. 

Consider a non-Gaussian spatial field with observations $y_i = y(\bs_i)$, $i = 1, \ldots, N$, recorded at spatial locations $\bs_1, \ldots, \bs_N$, where $N$ is typically large. To infer the non-Gaussian joint distribution of a spatial field from a small number of independent replicates, the ARGP approach utilizes a transport-map framework to factorize the joint distribution of $\by= (y_1, \ldots, y_N)^\top$ into a product of univariate conditional distributions:
\begin{equation}
    p(\by) = \prod_{i = 1}^N p(y_i |\by_{c_m(i)}),\label{eq:factorization}
\end{equation}
where $c_m(i)$ is a carefully chosen conditioning set of size at most $m$, with $m \ll N$ to ensure computational scalability. Specifically, $c_m(1) = \emptyset$, and $c_m(i) \subset \{1,\ldots,i-1\}$ for $i>1$. 

Crucial to the success and scalability of this approach is the ordering of the spatial locations and the subsequent selection of the conditioning sets. \citet{Katzfuss02042024} employs a \emph{maximin ordering} \citep[e.g.,][]{Guinness2016a,Schafer2020}, which begins with an arbitrarily selected initial location and sequentially chooses each subsequent location to maximize the minimum Euclidean distance to all previously ordered locations. Given this ordering, the conditioning set $c_m(i)$ is defined to contain the indices of the (up to) $m$ \emph{nearest spatial neighbors} of $\bs_i$ among the previously ordered locations. 

Let $l_i = \min_{j \in \{1,\ldots,i-1\}} \|\bs_i - \bs_j\|$ denote the distance from location $\bs_i$ to its nearest previously ordered neighbor. A key property of maximin ordering is that it first captures the field at a coarse global scale and then refines it at increasingly granular levels. Consequently, the length scale $l_i$ decays monotonically as $i$ increases. When $l_i$ decays, the conditional distributions in \eqref{eq:factorization} become increasingly Gaussian, even for highly non-Gaussian stochastic processes \citep{Katzfuss02042024}. 

By restricting the conditioning to these $m$ nearest previously ordered neighbors, the problem of inferring the $N$-variate distribution $p(\by)$ is transformed into $N$ independent lower-dimensional regressions of the form:
\begin{equation}
    y_i = f_i(\by_{c_m(i)}) + \epsilon_i, \quad \epsilon_i \sim \mathcal{N}(0,d_i^{2}), \quad i = 1, \ldots, N.\label{eq:regression}
\end{equation}

To estimate the unknown, potentially nonlinear functions $f_i$, a Bayesian nonparametric approach is utilized. Independent Gaussian-process and inverse-Gamma priors are assigned to each pair $(f_i,d_i^2)$:
\begin{equation}
        d_i^2 \stackrel{ind.}{\sim} \mathcal{IG}(\alpha_i,\beta_i), \qquad \text{with } \alpha_i>1, \; \beta_i > 0, \qquad i=1,\ldots,N, \label{eq:prior_ab}
\end{equation}
and
\begin{equation}
f_i | d_i \stackrel{ind.}{\sim} \GP(0,d_i^2 K_i), \qquad i=1,\ldots,N. \label{eq:f_d_GP}
\end{equation}
Here, $K_i$ is a covariance kernel defined as:
\begin{equation} \label{eq:K_i}
K_i(\mathbf{y}_{c_m(i)}, \mathbf{y}'_{c_m(i)})
= E(d_i^2)^{-1} \cdot \left(\mathbf{y}_{c_m(i)}^\top \mathbf{Q}_i \mathbf{y}'_{c_m(i)}
   + \sigma_i^2 \, \rho\left(\frac{\sqrt{\mathbf{y}_{c_m(i)}^\top \mathbf{Q}_i \mathbf{y}'_{c_m(i)}}}{\gamma}\right)\right),
\end{equation}
where $\rho$ is a correlation function. The spatial characteristics of the field are encoded into this kernel prior through the sparse precision matrix $\mathbf{Q}_i$ and the non-linear variance parameter $\sigma_i^2$, both of which are designed to vary with the decaying nearest-neighbor distance $l_i$. For common spatial covariances like the Mat\'ern kernel, the conditional variance $d_i^2$ also decays polynomially with the spatial scale $l_i$.

The parameters $\alpha_i,\beta_i,K_i,$ and $m$ are dictated by a small vector of global hyperparameters $\bftheta = (\theta_{d_1}, \theta_{d_2},\theta_{\sigma_1}, \theta_{\sigma_2}, \theta_{q}, \theta_{\gamma})$. The conditional density of $\by$ given $\bf$ and $\bd$ is the product of univariate normal densities:
\begin{equation}
p(\by |\bf, \bd) = \prod_{i=1}^N \normal(y_i | f_i(\by_{c_m(i)}),d_i^2).\label{eq:model}
\end{equation}

Now suppose that inference is based on training data $\bY = (\by^{(1)},\ldots,\by^{(n)})$, consisting of $n$ independent replicates from \eqref{eq:model}. The hyperparameters $\bftheta$ can be efficiently estimated by maximizing the integrated log-likelihood, $\hat\bftheta = \argmax_{\bftheta} \log p(\bY)$, via mini-batch gradient ascent, where $p(\bY)$ can be written as a product of $N$ closed-form terms. Because of the sparsity induced by the conditional structure introduced in \eqref{eq:factorization}, the computational complexity of evaluating each product term in the likelihood is drastically reduced to $\mathcal{O} (n^3 + mn^2)$. Finally, given the training data $\bY$ and the estimated hyperparameters, the posterior predictive distribution takes the form of a product of Student's $t$-distributions, which can be readily used to generate new conditional samples.

\rev{The spatial ARGP framework has since been extended to multivariate fields \citep{wiemann2023}, to improved data efficiency via parametric shrinkage \citep{chakraborty2025} and scalable composite transformations \citep{brachem2026dataefficientgenerativemodelingnongaussian}, and to multiple resolutions for downscaling \citep{Calle-Saldarriaga2025}. All of these approaches index the field by spatial coordinates alone and treat it as exchangeable across replicates; none models temporal dependence or supports temporal forecasting from a partially observed trajectory, which is the gap we address here.}

\section{Methodology: Spatio-temporal autoregressive Gaussian processes}\label{sec:methodology}

Consider a spatio-temporal field $\by = (y_{1},\ldots, y_N)^\top$, where each $y_{i}$ is observed at a spatio-temporal coordinate $\bx_i = (\bs_i, t_i)$, with spatial location $\bs_i$ (typically in 2 or 3 dimensions) and time point $t_i$. The time points $t_i$ may or may not be unique across observations. In many applications, data are collected on a fixed spatial grid of $N_s$ locations across $N_t$ time points, yielding $N = N_s \times N_t$. However, our proposed method is fully flexible and remains applicable even if the observed spatial locations vary over time, or if all observations are recorded at completely distinct continuous time points (i.e., $t_i \neq t_j$ for all $i \neq j$).

Our objective is to develop a scalable and flexible \rev{stochastic} model for the joint distribution of $\by$ based on a small number, denoted by $n$, of independent replicates, $\bY = (\by^{(1)}, \ldots, \by^{(n)})$. To this end, we extend the spatial autoregressive Gaussian process (ARGP) framework reviewed in Section~\ref{sec:review} to the spatio-temporal setting. The key challenge is to incorporate temporal structure into the autoregressive factorization in a way that preserves scalability to very high dimensions and that also enables conditional forecasting. Our approach addresses these challenges through a data-driven construction of space-time orderings and sparse conditioning sets, which we describe next.

\subsection{Ordering and conditioning sets} \label{sec:ord_and_cond}

The ARGP model described in Section \ref{sec:review} requires an ordering of the observations $y_1,\ldots,y_N$ and the selection of sparse conditioning sets $c_m(2),\ldots,c_m(N)$. For observations $y_{i} = y(\bx_{i})$, these tasks are carried out based on a distance metric between the inputs $\bx_{i}$. 

In purely spatial fields, Euclidean distance is standard. However, spatio-temporal fields are indexed by both space and time. To place spatial and temporal coordinates on a comparable scale, we define scaled coordinates $\tilde\bx_{i} = (\bs_i/\lambda_s, t_i/\lambda_t)$, where $\lambda_s$ and $\lambda_t$ are spatial and temporal correlation length-scale parameters estimated from the data (see Section~\ref{sec: estimation}). Their ratio determines the relative weighting of spatial and temporal separation. We then define the distance $r_{ij}$ between two scaled coordinates $\tilde\bx_i$ and $\tilde\bx_j$ as
\begin{equation}
    r_{ij}^2 = \|\tilde \bx_i - \tilde \bx_j\|^2 = \frac{\|\bs_i - \bs_j\|^2}{\lambda_s^2} + \frac{\|t_i - t_j\|^2}{\lambda_t^2}  =  \frac{1}{\lambda_s^2} \Big( \|\bs_i - \bs_j\|^2 + \eta\|t_i - t_j\|^2 \Big), \label{eq:distance}
\end{equation}
where $\eta = \lambda_s^2 / \lambda_t^2$ controls the relative scaling of time versus space. Given these scaled locations, we consider two strategies for constructing the ordering and conditioning sets.

\begin{figure}[tbp]
    \centering
    
    \begin{subfigure}[b]{\textwidth}
        \centering
        \makebox[0.41\textwidth][c]{\footnotesize\textbf{$i=9$}} \hfill
        \makebox[0.41\textwidth][c]{\footnotesize\textbf{$i=39$}}
        \vspace{2pt}
        
        \begin{subfigure}[b]{0.41\textwidth}
            \includegraphics[width=\textwidth]{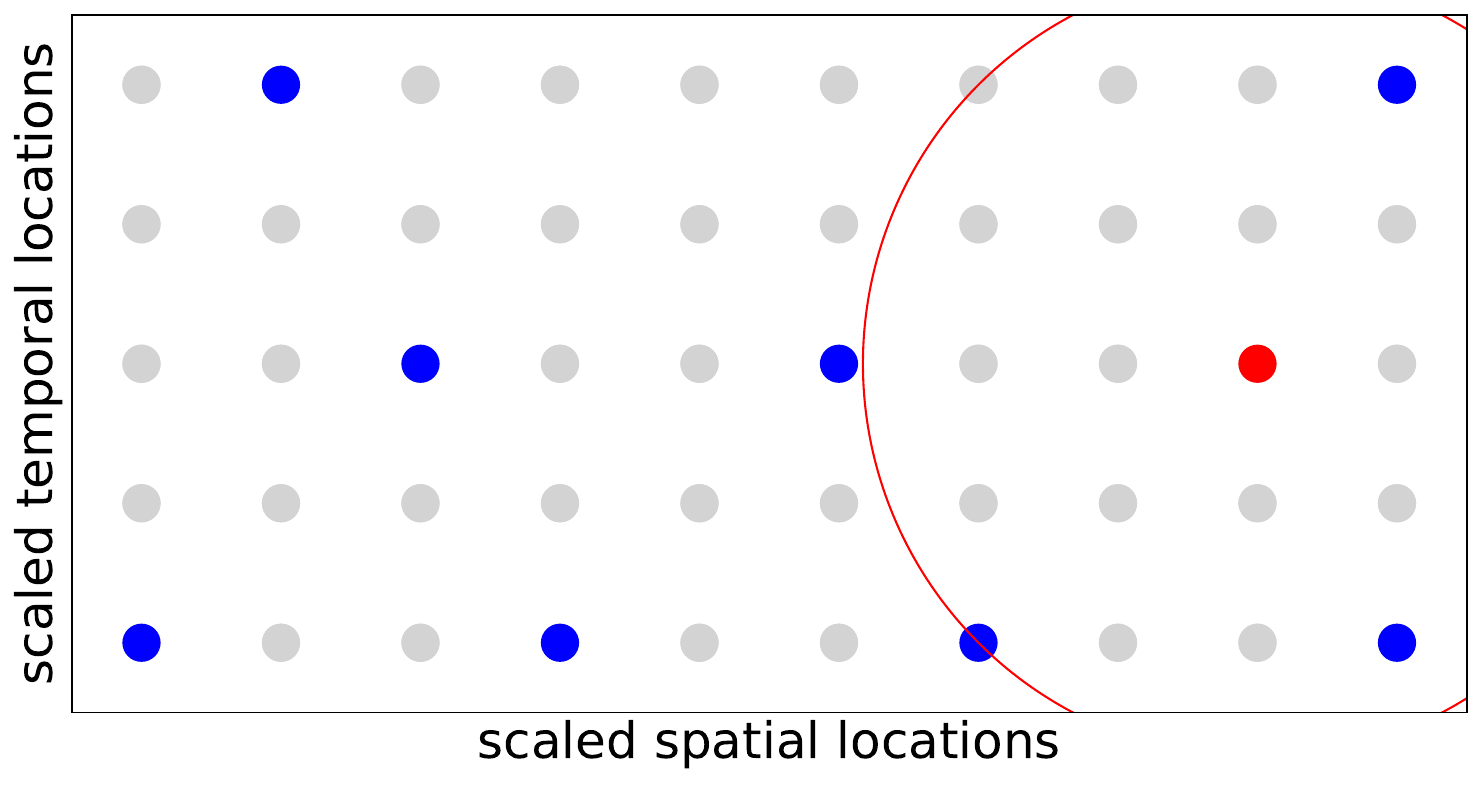}
        \end{subfigure}
        \hfill
        \begin{subfigure}[b]{0.41\textwidth}
            \includegraphics[width=\textwidth]{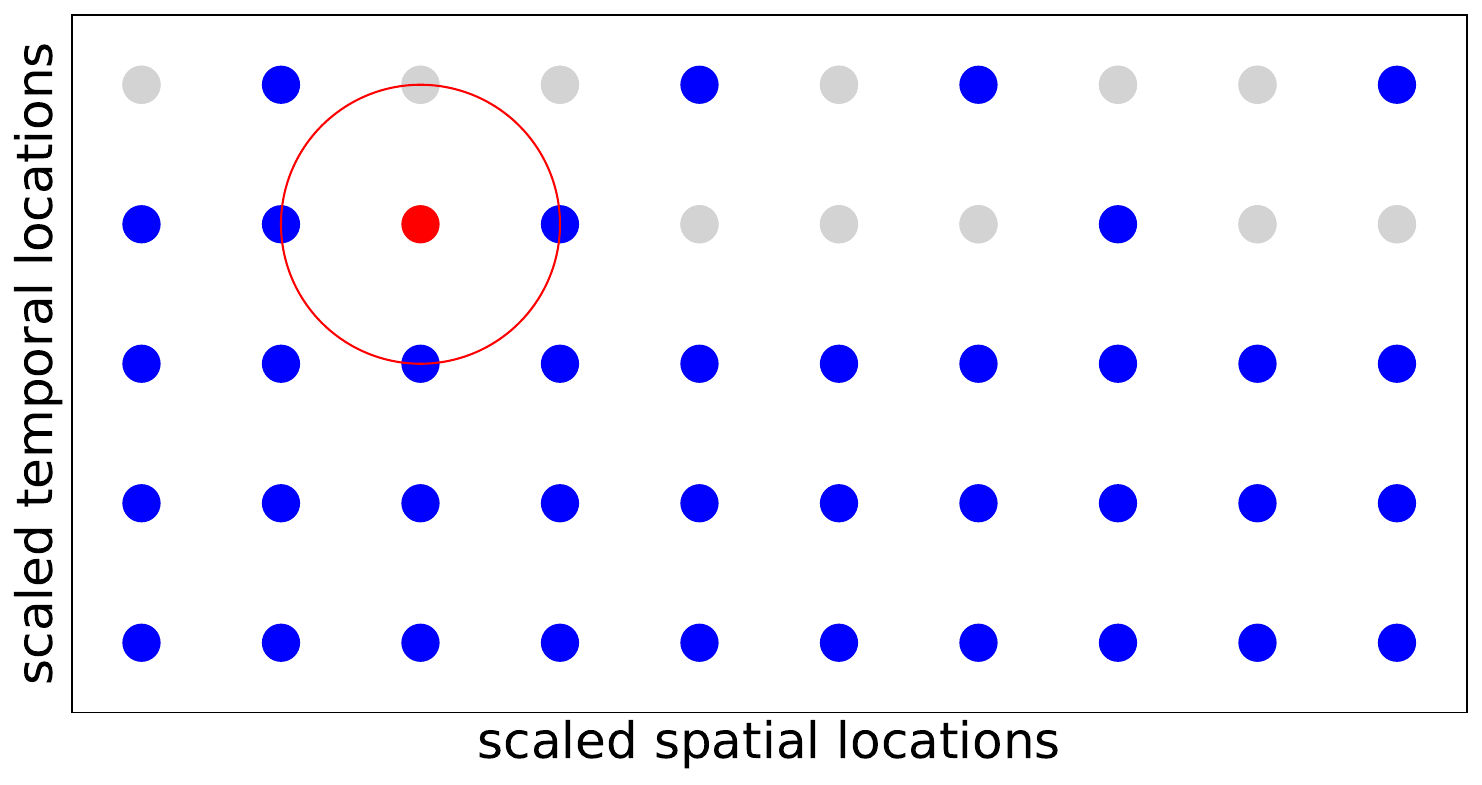}
        \end{subfigure}
        
        \vspace{5pt}
        
        \begin{subfigure}[b]{0.41\textwidth}
            \includegraphics[width=\textwidth]{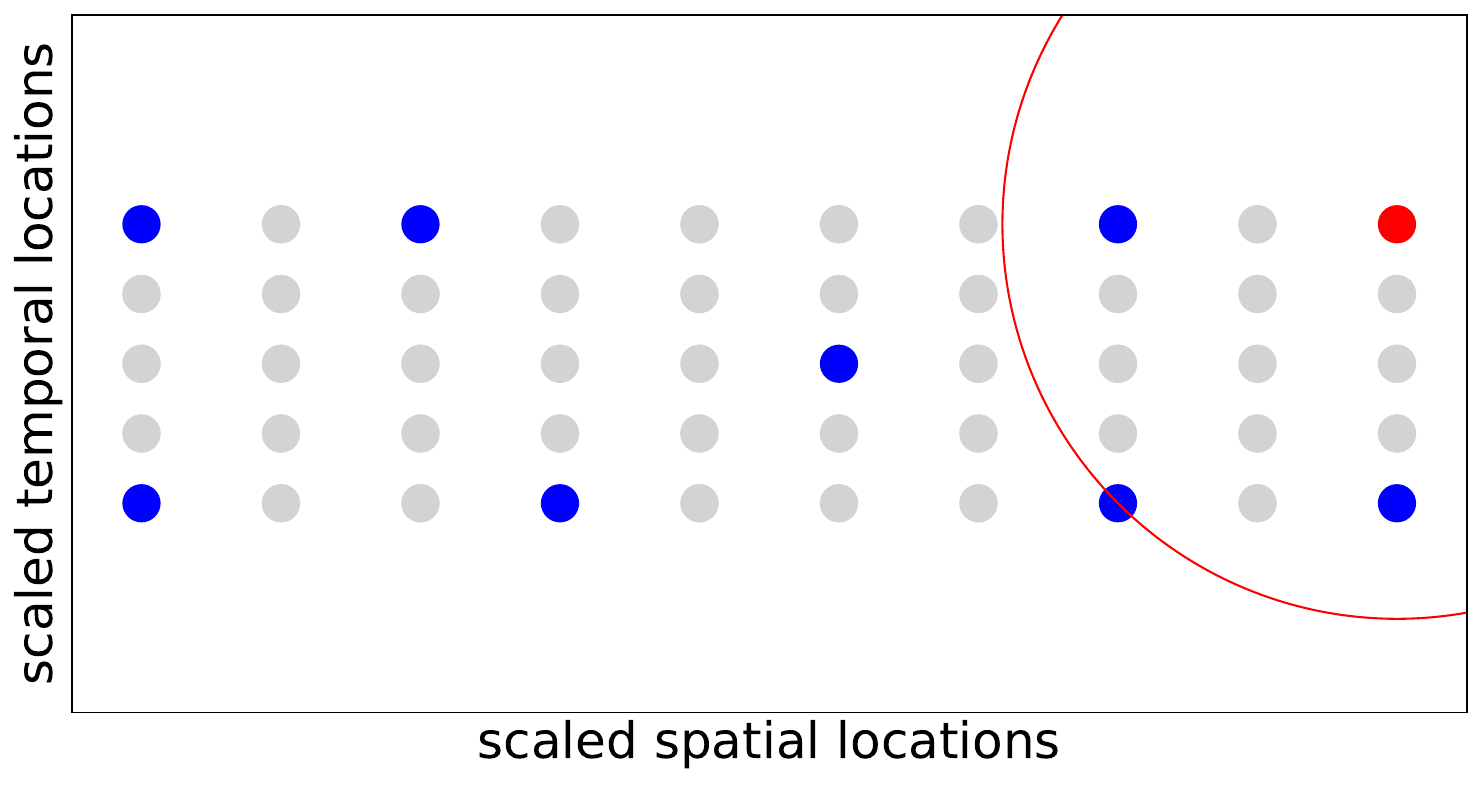}
        \end{subfigure}
        \hfill
        \begin{subfigure}[b]{0.41\textwidth}
            \includegraphics[width=\textwidth]{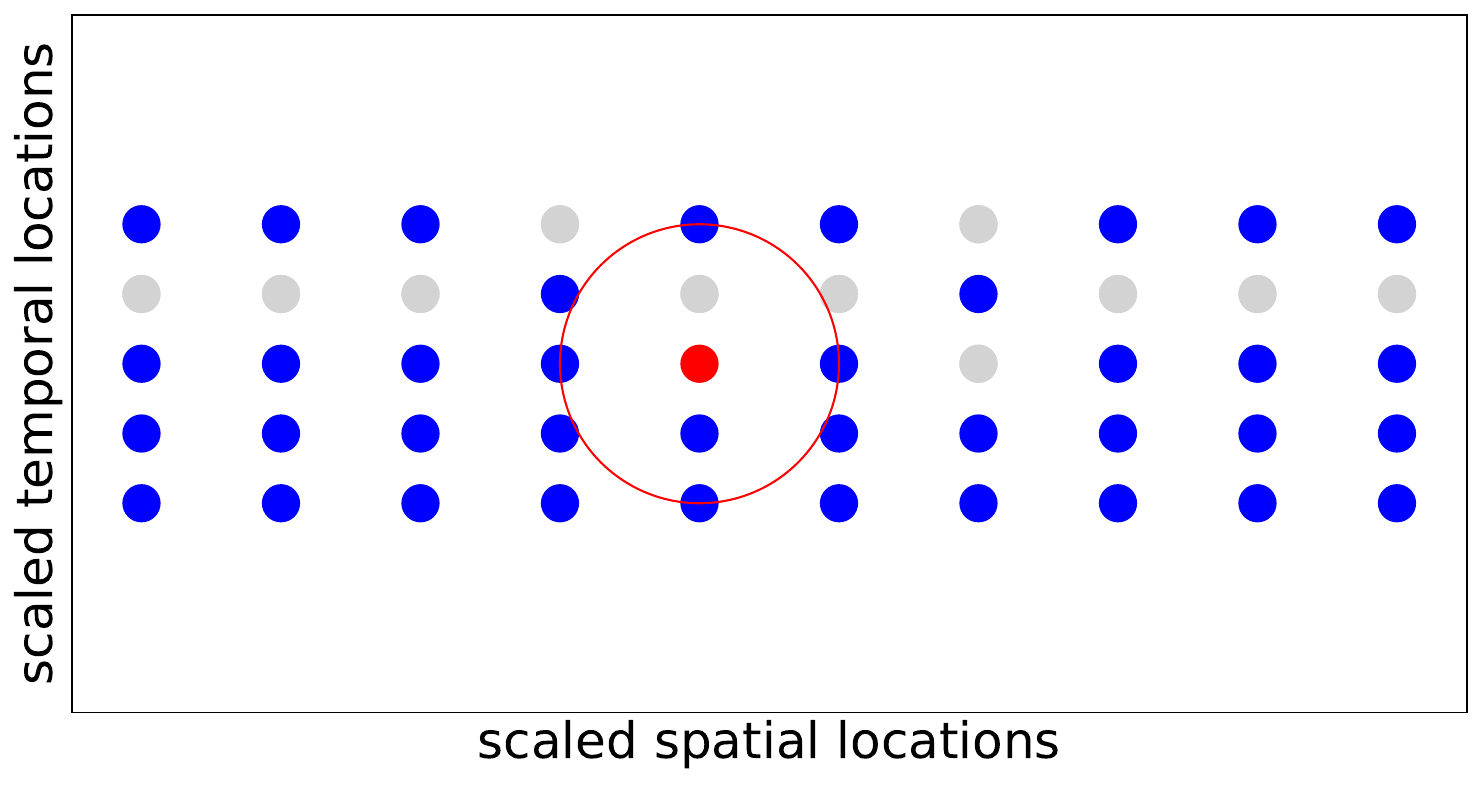}
        \end{subfigure}
        
        \captionsetup{font=small} 
        \caption{Illustration of the maximin ordering method for space-time data}
    \label{fig:maximin_ordering}
    \end{subfigure}

    \vspace{10pt} 
    \begin{subfigure}[b]{\textwidth}
        \centering
        \makebox[0.41\textwidth][c]{\footnotesize\textbf{$i=9$}} \hfill
        \makebox[0.41\textwidth][c]{\footnotesize\textbf{$i=39$}}
        \vspace{2pt}

        \begin{subfigure}[b]{0.41\textwidth}
            \includegraphics[width=\textwidth]{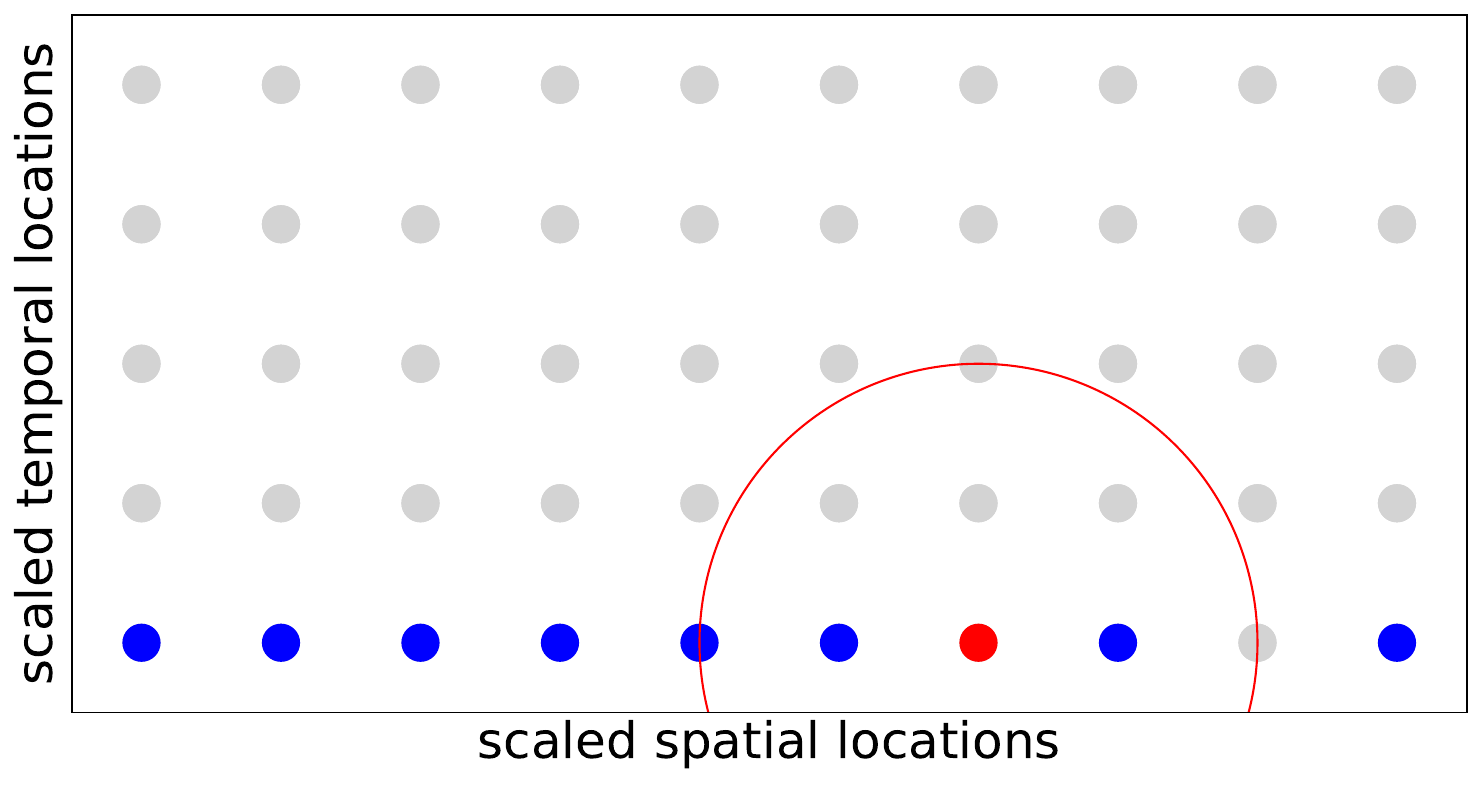}
        \end{subfigure}
        \hfill
        \begin{subfigure}[b]{0.41\textwidth}
            \includegraphics[width=\textwidth]{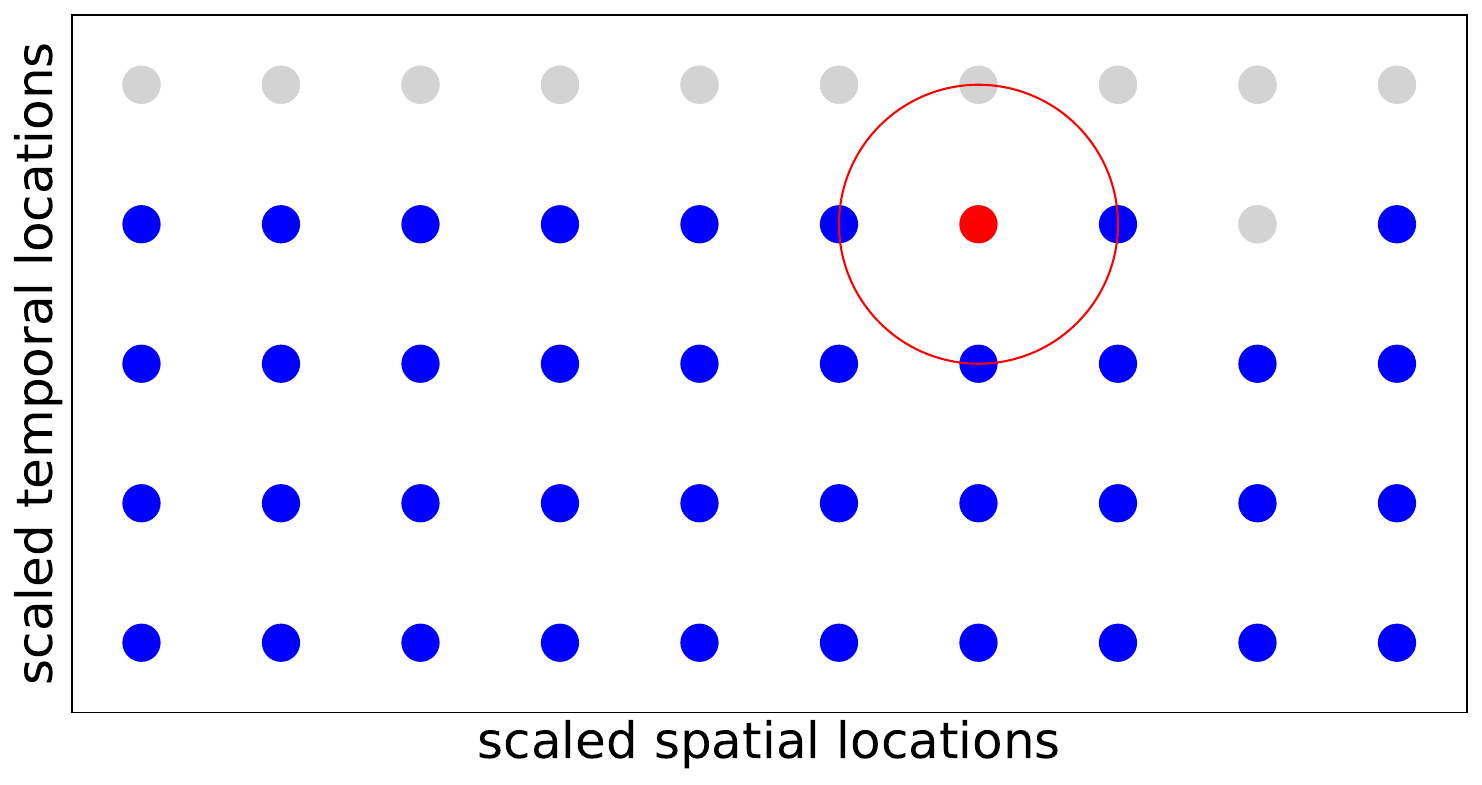}
        \end{subfigure}
        
        \vspace{5pt}
        
        \begin{subfigure}[b]{0.41\textwidth}
            \includegraphics[width=\textwidth]{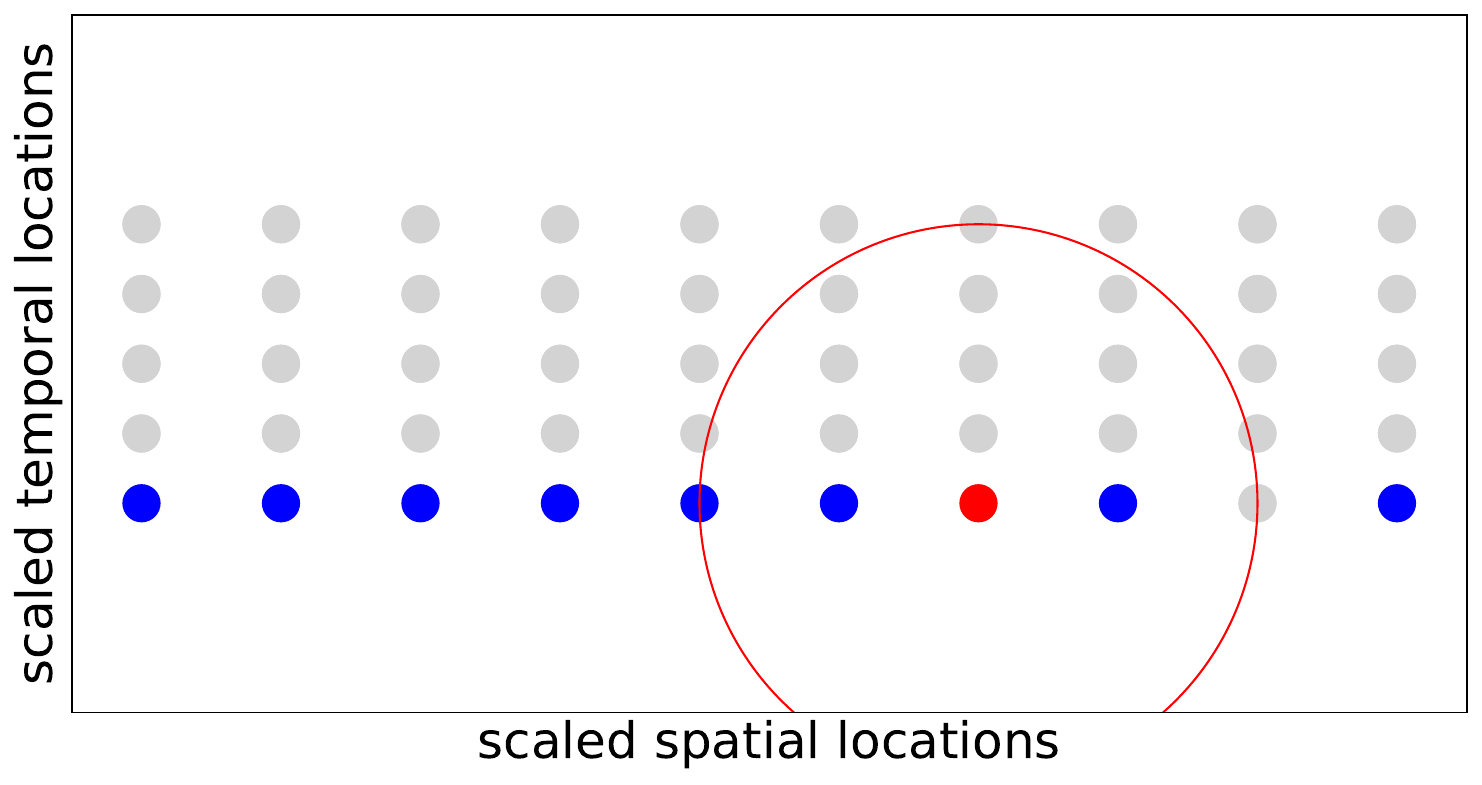}
        \end{subfigure}
        \hfill
        \begin{subfigure}[b]{0.41\textwidth}
            \includegraphics[width=\textwidth]{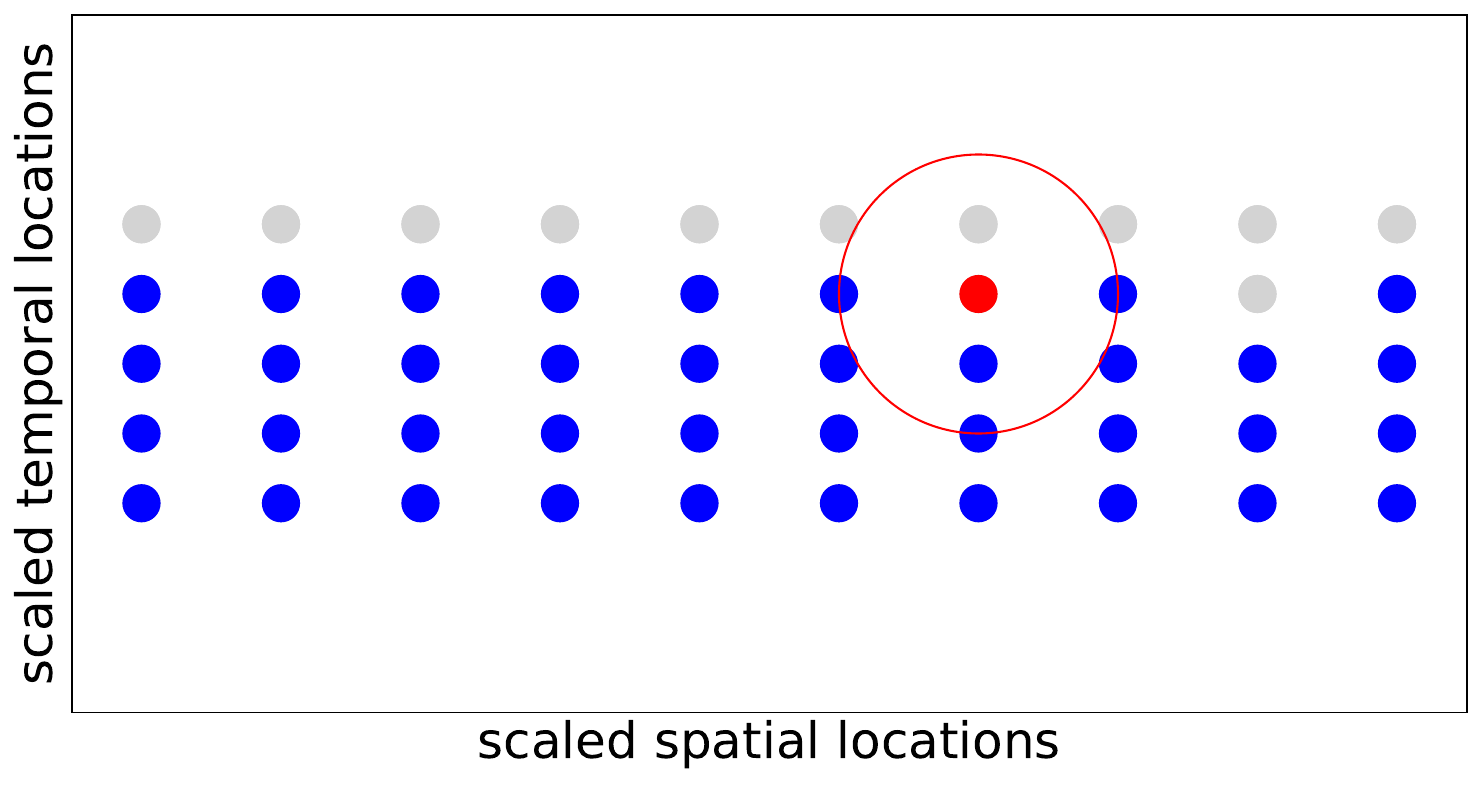}
        \end{subfigure}

        \captionsetup{font=small}
        \caption{Illustration of the time ordering method for space-time data}
    \label{fig:time_ordering}
    \end{subfigure}

    \caption{\textbf{Maximin (a) and time (b) ordering in a space--time domain.} Rows show weaker (top) and stronger (bottom) temporal dependence. The $i$-th location is red, ordered points are blue, and unordered points are gray. Circles represent the conditioning radius for $c_m(i)$ ($m=3$). Columns correspond to early ($i=9$) and later ($i=39$) ordering stages.}

\end{figure}

\paragraph{Maximin space-time ordering.} We can apply the standard maximin ordering \citep{Katzfuss2017a, Katzfuss02042024} directly to the scaled space-time domain. In this approach, time is treated simply as an additional dimension to the spatial domain. The algorithm arbitrarily selects the first spatio-temporal location and sequentially selects each subsequent location to maximize the minimum distance to all previously ordered locations. Formally, this produces a sequence of indices $\mathcal{M} = \mathcal{M}(\mathbf{\tilde{X}})$ based on the scaled coordinates $\mathbf{\tilde{X}}$, such that the $i$-th element is:
\begin{equation} 
    \mathcal{M}_i = \argmax_{j \in \{1, \ldots, N\} \setminus \mathcal{M}_{1: i-1}} \min_{k \in \mathcal{M}_{1:i-1}} \|\tilde\bx_k - \tilde\bx_j\|. \label{eq:maxmin}
\end{equation} 
After reordering, the conditioning set $c_m(i)$ is constructed from the indices of the $m$ nearest neighbors of $\tilde\bx_i$ among the previously ordered points $\tilde\bx_1,\ldots,\tilde\bx_{i-1}$. For a broad class of processes, maximin ordering results in an exponential decay of influence as a function of neighbor distance in the univariate conditional distributions in \eqref{eq:factorization}; this ensures that an accurate approximation can be achieved using a very small number of neighbors $m$ \citep{Schafer2020}. Figure~\ref{fig:maximin_ordering} illustrates the resulting sequence of scaled spatio-temporal locations under this ordering. At an early stage of the ordering ($i=9$), the conditioning set draws neighbors from different (earlier and later) time frames, reflecting the limited number of previously ordered points. As the ordering progresses ($i=39$), the conditioning set increasingly favors spatial neighbors within the same time frame, particularly when temporal dependence is weak.
Another useful property of the maximin ordering is that it ensures that points early in the ordering are spread maximally throughout the space-time domain; by storing only these early points, a given space-time field can be compressed and approximately reconstructed in what can be viewed as a nonlinear spatio-temporal version of principal component analysis, as illustrated for spatial fields in \citet[][App.~F]{Katzfuss02042024}. \rev{Maximin ordering has also been found to compare favorably to alternative ordering
schemes in Gaussian space-time settings \citep{idir2025improvingspatiotemporalgaussianprocess}, although that study also reports a random ordering to perform comparably while requiring no space-time distance.
We retain maximin ordering because it does more here than control approximation accuracy.
The prior is itself indexed by the nearest-neighbor distance $l_i$ (Section \ref{sec:parameterization}), and the coarse-to-fine refinement produced by maximin ordering is what makes the univariate conditionals in \eqref{eq:factorization} increasingly Gaussian as $i$ grows, which is the behavior our Gaussian-process prior is constructed to exploit.
A random ordering would also forfeit the compression property just described.}

\paragraph{Time ordering.} Alternatively, we can use time as the primary sorting key. This aligns naturally with the chronological progression of data collection and is particularly useful for forecasting \rev{\citep{gyger2026scalablenonseparablespatiotemporalgaussian}}. In this scheme, observations are first ordered by time, so that $t_i \le t_j$ whenever $i < j$. If multiple spatial responses share the exact same temporal coordinate, they are subsequently ordered using the spatial maximin ordering based on their spatial coordinates $\bs$. 

Mathematically, time ordering produces a sequence of indices $\bT(\tilde{\bX})$ defined as:
\begin{equation} 
\bT(\tilde{\bX}) = \Big(\mathcal{M}\big(\mathcal{S}(\tilde{t}_{(1)})\big), \; k_1 + \mathcal{M}\big(\mathcal{S}(\tilde{t}_{(2)})\big), \; \ldots, \; k_{N_t - 1} + \mathcal{M}\big(\mathcal{S}(\tilde{t}_{(N_t)})\big)\Big),
\end{equation} 
where $\mathcal{S}(t_i) = \{\bs : (\bs, t_i) \in \{\bx_{i}\}_{i = 1}^{N}\}$ is the set of points at time $t_i$, $k_i = \sum_{j=1}^{i - 1}|\mathcal{S}(t_{(j)})|$, $N_t$ is the number of unique temporal coordinates, and $\mathcal{M}$ is the spatial maximin sequence from \eqref{eq:maxmin}. The conditioning sets are still chosen as the $m$ nearest neighbors in the scaled spatio-temporal domain among previously ordered observations. Algorithm \ref{alg:time_ordering} summarizes this procedure, and Figure \ref{fig:time_ordering} illustrates the resulting conditioning-set structures, where the responses are restricted to condition only on historical responses (i.e., responses from preceding time points). Early in the ordering ($i=9$), the conditioning set includes neighbors from the same time frames, reflecting the limited number of previously ordered points. As the ordering progresses ($i=39$), time ordering also selects neighbors that are close in space but drawn from earlier time steps, leading to conditioning sets that span across time. This property of conditional sets selected under time ordering is leveraged in our method for forecasting simulation.

Throughout the remainder of the paper, we assume that the data are ordered using one of these two schemes.

\begin{algorithm}[H]
\DontPrintSemicolon
\caption{Time Ordering with Maximin Tie-Breaking}
\label{alg:time_ordering}
\KwIn{$\bX \in \mathbb{R}^{N \times (d+1)}$ }
\KwOut{$\br$: order of indices, a vector of length $N$}
$\bt \gets \mbox{sort}(\bX[:, d+1])$\;
$\bt^* \gets \text{unique}(\bt)$ \tcp*[r]{$\bt^*$ contains the unique temporal entries} 
Initialize $\br$ as a vector of length $N$, $\mbox{offset} \gets 0$ \;
\For{$t \in \bt^*$} {
    $\mbox{ind}_t \gets \mbox{which}(\bX[:, d+1] == t)$, $N_t \gets |\mbox{ind}_t|$, $\bX_t \gets \bX[\mbox{ind}_t, 1:d]$ \;
    $\mbox{ind}_{mm} \gets \mbox{maximin}(\bX_t)$ \;
    $\br[\mbox{offset} + 1 : \mbox{offset} + N_t] = \mbox{ind}_t[\mbox{ind}_{mm}]$ \;
    $\mbox{offset} \gets \mbox{offset} + N_t$\;
}
\Return $\br$\;
\end{algorithm}

\FloatBarrier

\subsection{Estimation of the space-time scaling\label{sec: estimation}}

This scaled geometry is central to the sparsity in ARGP, because it determines which observations are most informative for each conditional factor and thereby controls both statistical efficiency and computational sparsity. The selected nearest neighbors depend on the relative scaling of the spatial and temporal dimensions (see \eqref{eq:distance}). To avoid distortions caused by heterogeneous units across coordinate dimensions, we first normalize all spatial dimensions to a common scale, allowing us to attribute metric variations solely to the relative strength of the temporal coordinate.
We cannot estimate the scale parameters $\lambda_s$ and $\lambda_t$ simultaneously with the ARGP transport-map parameters using gradient-based optimization, because the ordering and the conditioning sets $c_m(i)$ depend on $\lambda_s$ and $\lambda_t$, the change to which causes discrete, non-differentiable changes in the neighbor sets. 

Therefore, we obtain $\lambda_s$ and $\lambda_t$ in an initial-stage learning. Specifically, we fit a GP with an isotropic Mat\'ern kernel (smoothness $\nu=1.5$) in the scaled input space. At this stage, we aim to obtain estimates of $\lambda_s$ and $\lambda_t$ to approximate the relative strength of the temporal coordinate for the sole purpose of ARGP ordering and conditioning-set selection. Hence, the GP is trained on a randomly selected subset of the training data for computational speed. We optimize the scale parameters by minimizing the negative log-likelihood using the PyTorch automatic-differentiation framework. To guarantee robustness, we repeat this estimation five times across different randomly selected subsets and average the resulting length scales. In our experiments, the standard errors of the estimated scale parameters were negligible, indicating highly consistent estimates across different subsets. This two-stage procedure is summarized in Algorithm \ref{alg:length_scales}.

\begin{algorithm}[H]
\DontPrintSemicolon
\caption{Estimate length scales via a parametric Gaussian process}
\label{alg:length_scales}
\KwIn{\rev{$\bX \in \mathbb{R}^{N \times (d+1)}, \bY \in \mathbb{R}^{n \times N}$; $N_{\mathrm{samp}}$, $n_{\mathrm{samp}}$, $n_{\mathrm{epoch}}$}}
\KwOut{$\lambda_s, \lambda_t$}
\BlankLine
Sample $S_N \subset \{1, \dots, N\}$ and $S_n \subset \{1, \dots, n\}$ s.t.\ $|S_N| = N_{\mathrm{samp}}$ and $|S_n| = n_{\mathrm{samp}}$\;
\BlankLine
\BlankLine
\rev{$\bX_s \gets \bX[S_N, :]$, \quad $\bY_s \gets \bY[S_n, S_N]^\top$} \tcp*{\rev{one column of $\bY_s$ per replicate}}
\BlankLine
\BlankLine
Initialize $\mathcal{K}$ to be Mat\'ern covariance kernel with smoothness 1.5
\BlankLine
\For{$k \gets 1$ \KwTo $n_{\mathrm{epoch}}$}{
    \rev{$\tilde\bX_s \gets \big[\, \bX_s[:, 1:d]/\lambda_s, \;\; \bX_s[:, d+1]/\lambda_t \,\big]$} \tcp*{\rev{rescale the original coordinates}}
    \rev{$\bK \gets \mathcal{K}(\tilde\bX_s)$}  \tcp*{Compute covariance matrix}
    \rev{$nll \gets \frac{1}{2}\tr(\bY_s^\top \bK^{-1}\bY_s) + \frac{n_{\mathrm{samp}}}{2}\log\det \bK$}\;
    Update $\mathcal{K}$ hyperparameters and $(\lambda_s, \lambda_t)$ by minimizing $nll$ via gradient descent\;
}
\BlankLine
\Return{$\lambda_s, \lambda_t$}\;
\end{algorithm}

\subsection{Parameterization of the prior distributions} \label{sec:parameterization}

\begin{figure}[ht]
\centering

\begin{tabular}{c c c}
     \textbf{Maximin} & \textbf{Time} \\[-0.5em]

    \begin{subfigure}{0.45\textwidth}
        \centering
        \includegraphics[width=\textwidth]{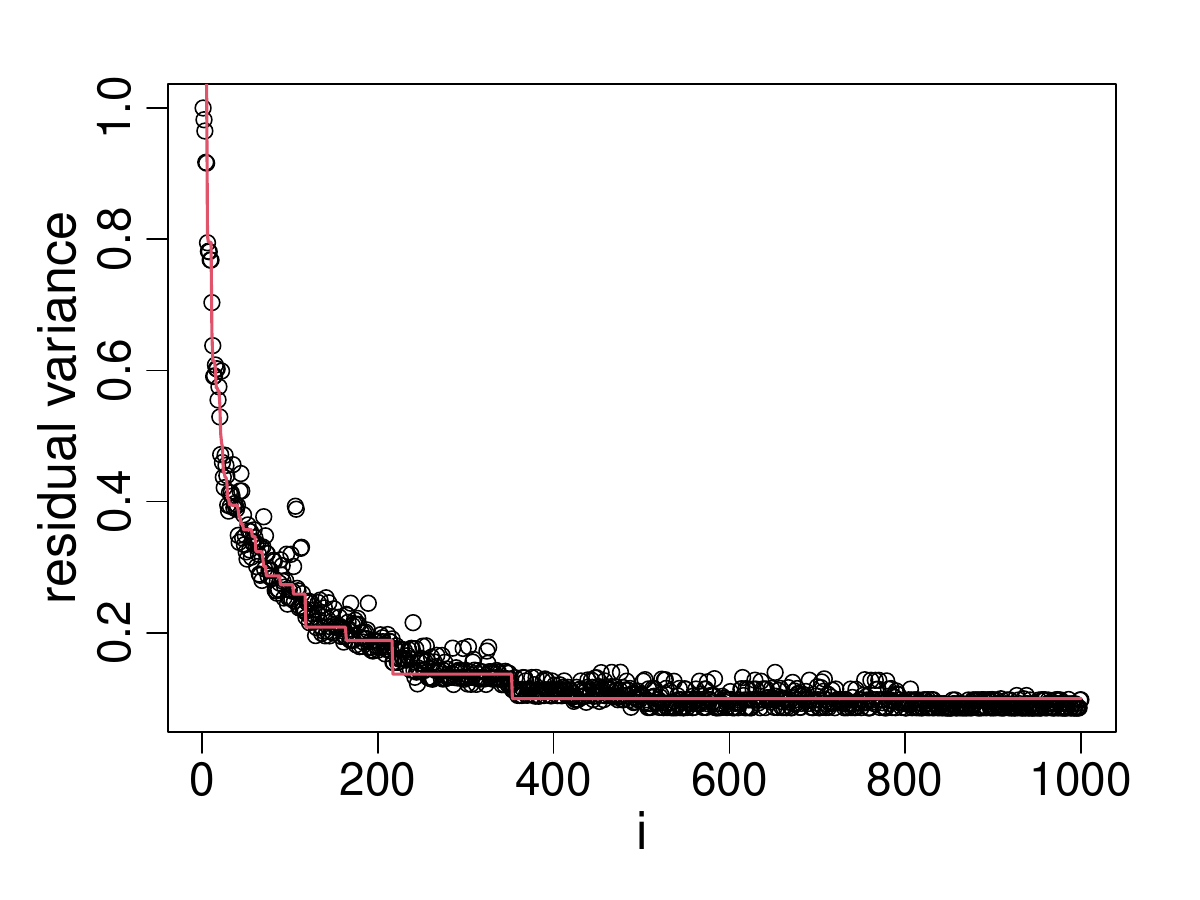}
        \caption{}
        \label{fig:di_sp03}
    \end{subfigure}
    &
    \begin{subfigure}{0.45\textwidth}
        \centering
        \includegraphics[width=\textwidth]{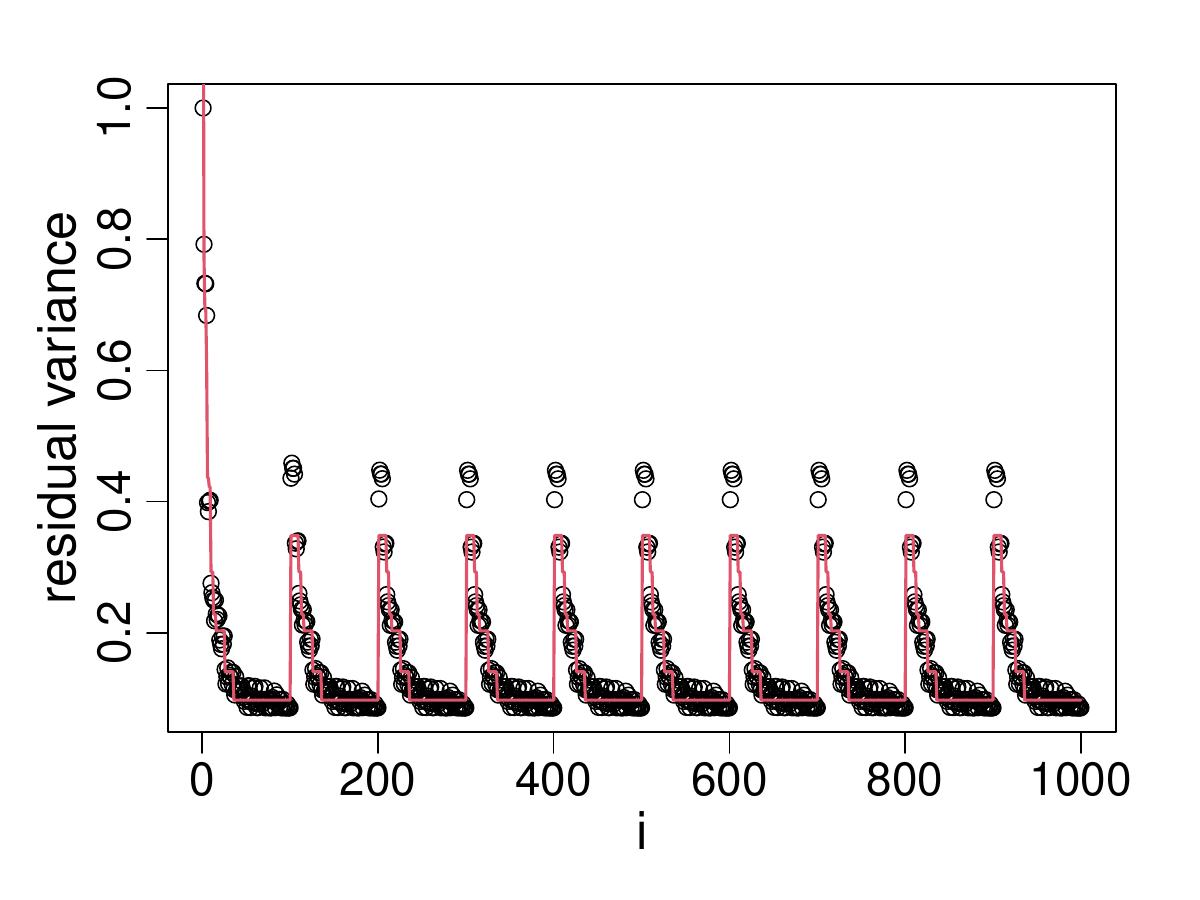}
        \caption{}
        \label{fig:di_sp03_time}
    \end{subfigure}
    \\[0em]

    \begin{subfigure}{0.45\textwidth}
        \centering
        \includegraphics[width=\textwidth]{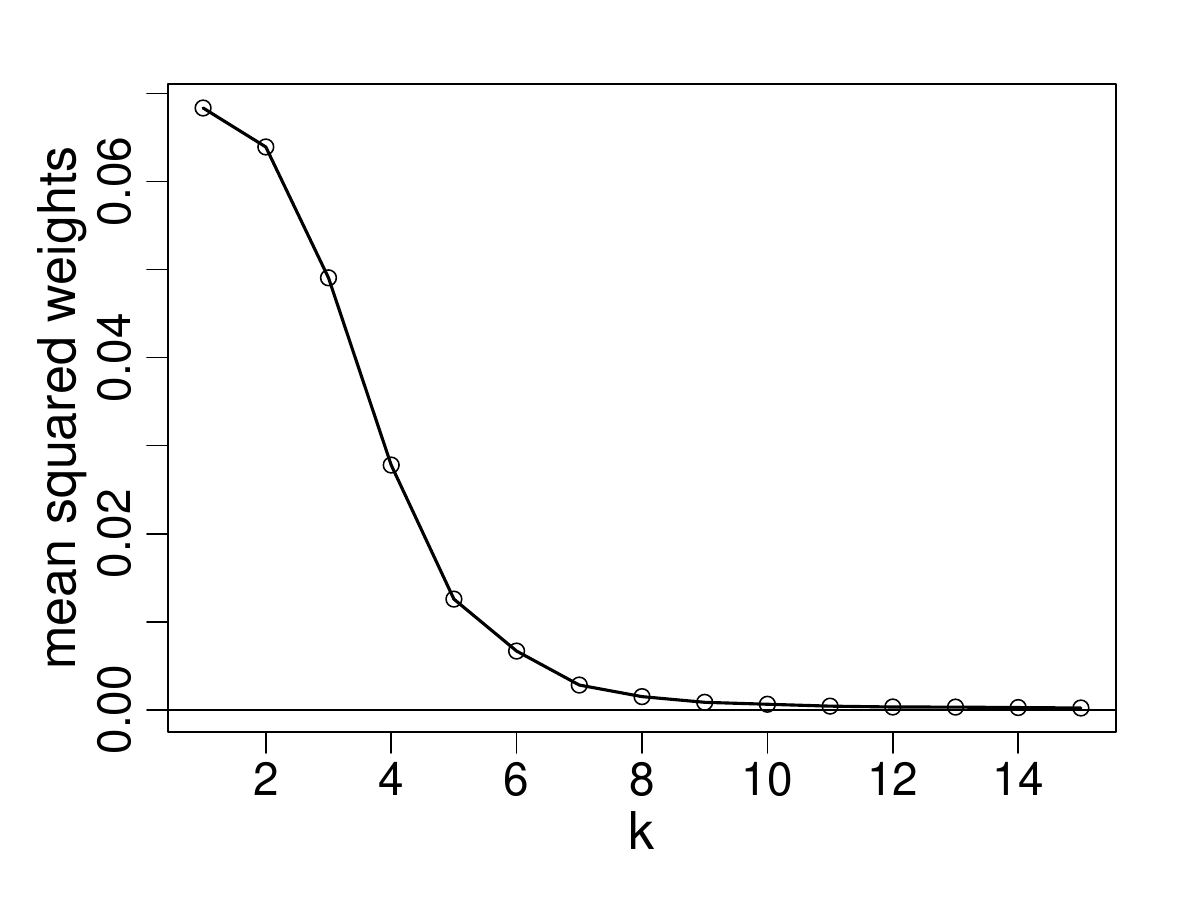}
        \caption{}
        \label{fig:k_sp03_maximin}
    \end{subfigure}
    &
    \begin{subfigure}{0.45\textwidth}
        \centering
        \includegraphics[width=\textwidth]{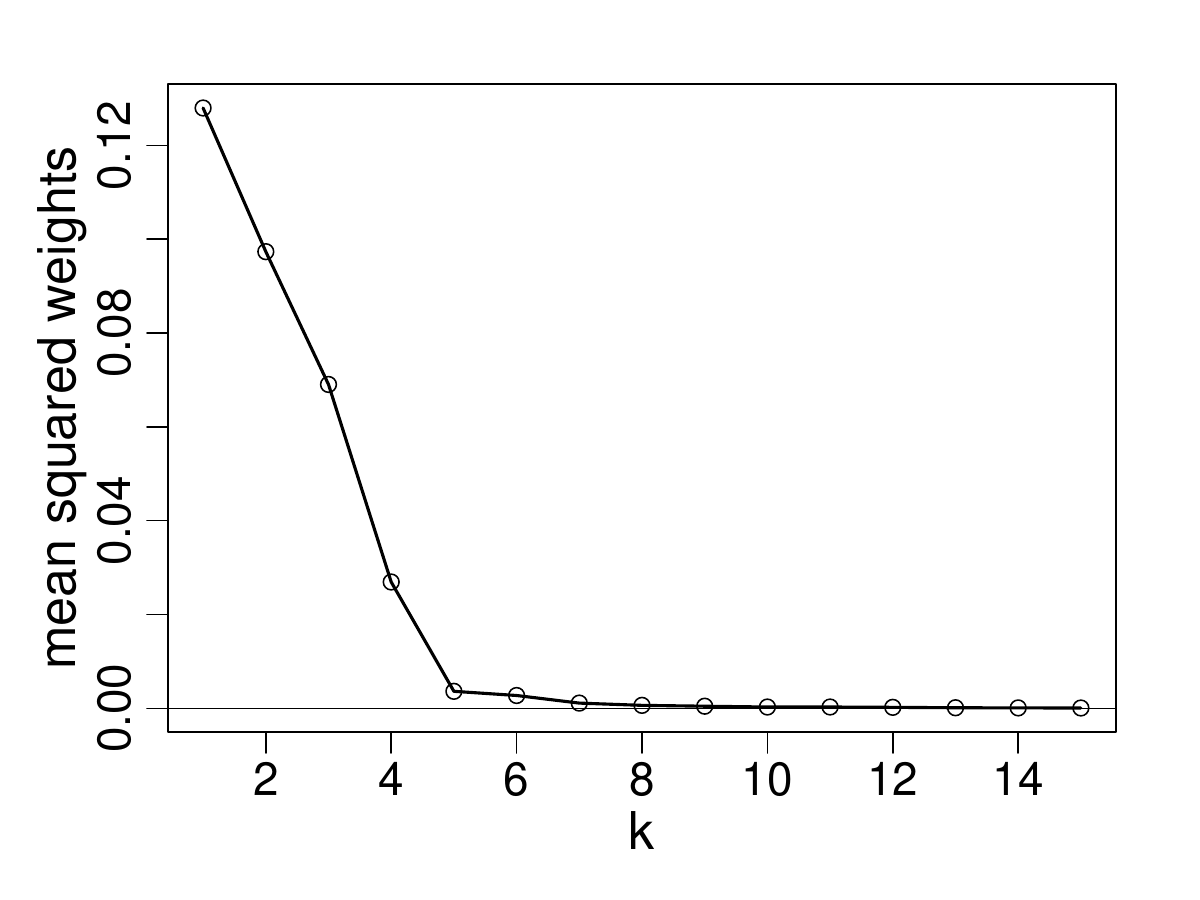}
        \caption{}
        \label{fig:k_sp03_time}
    \end{subfigure}
    \\[-0.5em]
\end{tabular}

\caption{\textbf{Decay relationships under maximin ordering (left) and time ordering (right).} 
Results are from a Gaussian field on a regular spatio-temporal grid ($N=10^3$) with a Mat\'ern covariance ($\nu=0.3$, $\rho=0.5$). 
Top row: Conditional variance ($d_i^2 \ (\circ)\ \text{and}\ e^{\theta_{d,1}} \, \ell_i^{\theta_{d,2}} \ (\textcolor{red}{-})$). 
Bottom row: Average squared regression coefficients ($b_{i,k}^2, \text{over } i = 1, \ldots, N$ for a fixed $k$). 
Note the periodic jumps in the time-ordering panel due to transitions between discrete time frames.}
\label{fig:decays}
\end{figure}

To make our method feasible in high-dimensional settings, we parameterize the prior distributions for the conditional variance $d_i^2$ and the conditional mean function $f_i$ by exploiting distance decay behaviors, similar to \rev{\citet{Katzfuss02042024}}. \citet{Schafer2020} demonstrated that for Gaussian processes with covariance functions equivalent to Green's functions of elliptic PDEs of order $r$, the conditional variance $d_i^2$ decays as $l_i^{2r}$. 

\paragraph{Prior on conditional variance $d_i^2$} 
Motivated by \rev{this $l_i^{2r}$ decay result of \citet{Schafer2020}}, we parameterize the expected conditional variance as
$$E(d_i^2) = \exp(\theta_{d,1} + \exp(\theta_{d,2}) \log(l_i)),$$
where $\theta_{d,1}$ and $\theta_{d,2}$ are hyperparameters to be learned from data. We assume the standard deviation of the inverse-Gamma prior is proportional to its expectation via a fixed scaling factor $g$, which completes the prior specification. As illustrated in Figure \ref{fig:di_sp03}, this parameterization captures the polynomial decay well under maximin space-time ordering. 

Under time ordering, however, we observe a periodic decay pattern when coordinates fall on a regular temporal grid (Figure \ref{fig:di_sp03_time}). This phenomenon occurs because the first few locations queried in a newly observed time frame must draw their nearest neighbors from the preceding time frame (which are further away). As more coordinates within the current time frame are subsequently included, nearest neighbors are selected from the same time frame, and the minimum distance towards the nearest neighbors $l_i$ drops sharply before stabilizing. The amplitude of these periodic `jumps' depends on the strength of the temporal correlation. Nonetheless, under both weak and strong temporal correlation scenarios, our parameterization of $E(d_i^2)$ provides a stable and accurate fit to the decay envelope.

\paragraph{Priors on mean functions $f_i $} 
The mean function is governed by the diagonal precision matrix $\mathbf{Q}_i = \operatorname{diag}(q_{i,1}^2, \ldots, q_{i,i-1}^2)$. A large $q_{i,k}$ encapsulates the prior belief that the corresponding linear coefficient $b_{i,k}$ is close to zero, so that the $k$-th neighbor contributes little to inferring $y_i$. Motivated by the spatial \textit{screening effect} \citep{Stein2011}, we expect $y_i$ to depend primarily on its closest neighbors. Figures \ref{fig:k_sp03_maximin} and \ref{fig:k_sp03_time} verify that the squared empirical regression coefficients $b_{i,k}^2$ decay rapidly toward zero as the neighbor index $k$ increases, under both orderings. We thus parameterize the diagonal entries of $\mathbf{Q}_i$ via a single parameter $\theta_q$ as 
$$q_{i,k}^{2} = \exp(-k \exp(\theta_q)),$$ 
assuming that $\by_{c_m(i)}$ is ordered based on the ascending distance from $y_{i}$. Finally, to encourage the non-linearities predominantly at local scales, the variance of the non-linear component is set to decay at the same rate as the distance $l_i$: 
$$\sigma_i^2=\exp (\theta_{\sigma, 1}+\exp(\theta_{\sigma,2}) \log(l_i)),$$
and the range parameter $\gamma$ simply adopts an exponential transformation $\gamma = \exp(\theta_{\gamma})$. 

By assuming these structural decays, the massive number of local regression parameters ($3N + 2$) is reduced to a 6 (global) hyperparameters: $\bftheta= (\theta_{\sigma,1}, \theta_{\sigma,2}, \theta_{d,1}, \theta_{d,2}, \theta_{\gamma}, \theta_{q})$, enhancing the feasibility and efficiency in model training.

\subsection{Inference}

Due to the conjugate prior formulation, the marginal likelihood $p_{\boldsymbol{\theta}}(\mathbf{Y})$ has a tractable closed-form expression. This enables empirical Bayes inference, where the hyperparameters $\boldsymbol{\theta}$ are estimated by maximizing the log-likelihood using gradient-based optimization. Furthermore, stochastic optimization is also straightforward by selecting a subset of $\{1, \dots, N\}$ as one mini-batch. \rev{Note that the mini-batch is taken over the $N$ ordered responses, not over the $n$ replicates. At any fixed $\bftheta$, the integrated log-likelihood is, up to an additive constant, a sum of $N$ closed-form terms $L_i(\bftheta)$, one per ordered response. Drawing the mini-batch $B$ uniformly at random from $\{1,\ldots,N\}$ therefore makes $\frac{N}{|B|}\sum_{i \in B} \nabla L_i(\bftheta)$ an unbiased estimator of the full gradient.} As in \citet{Katzfuss02042024}, the integrated likelihood is given by:
\begin{equation}
p_{\boldsymbol{\theta}}(\mathbf{Y})
\propto 
\prod_{i=1}^{N} 
\left(
|\mathbf{G}_i|^{-1/2}
\cdot
\frac{\beta_i^{\alpha_i}}{\tilde{\beta}_i^{\tilde{\alpha}_i}}
\cdot
\frac{\Gamma(\tilde{\alpha}_i)}{\Gamma(\alpha_i)}
\right),
\end{equation}
where
\begin{equation}
\begin{aligned}
&\tilde{\alpha}_i = \alpha_i + \frac{n}{2}, \qquad
\tilde{\beta}_i  = \beta_i + \frac{1}{2}\mathbf{y}_i^{\top}\mathbf{G}_i^{-1}\mathbf{y}_i, \qquad
\mathbf{G}_i = \mathbf{K}_i + \mathbf{I}_n, \\[6pt]
&\mathbf{K}_i = K_i(\mathbf{Y}_{1:i-1},\, \mathbf{Y}_{1:i-1})
= \bigl( K_i(\mathbf{y}_{1:i-1}^{(j)},\, \mathbf{y}_{1:i-1}^{(l)}) \bigr)_{j,l=1,\ldots,n},\\
&\mathbf{Y}_{1:i-1} = [\mathbf{y}_{1:i-1}^{(1)}, \mathbf{y}_{1:i-1}^{(2)}, \ldots, \mathbf{y}_{1:i-1}^{(n)}]^{\top} \mbox{ and }\mathbf{y}_{1:i-1}^{(l)} \mbox{ is the $\mathbf{y}_{1:i-1}$ from the $l$th replicate.}
\end{aligned}
\end{equation}

Once $\hat{\boldsymbol{\theta}}$ is estimated, the posterior predictive distribution for a new sample $\mathbf{y}^\star$ can be expressed as a product of univariate Student's $t$-distributions:
\begin{equation}
p(\mathbf{y}^\star \mid  \hat{\boldsymbol{\theta}} )
=
\prod_{i=1}^{N}
t_{2\tilde{\alpha}_i}
\!\left(
y_i^\star \,\middle|\,
\hat{f}_i(\mathbf{y}_{1:i-1}^\star),\ 
\hat{d}_i^{\,2}\bigl( v_i(\mathbf{y}_{1:i-1}^\star) + 1 \bigr)
\right),
\end{equation}
where 
\begin{equation}
\begin{aligned}
v_i(\mathbf{y}_{1:i-1}^{\star}) 
&= 
K_i(\mathbf{y}_{1:i-1}^{\star}, \mathbf{y}_{1:i-1}^{\star})
-
K_i(\mathbf{y}_{1:i-1}^{\star}, \mathbf{Y}_{1:i-1})
\,\mathbf{G}_i^{-1}\,
K_i(\mathbf{Y}_{1:i-1}, \mathbf{y}_{1:i-1}^{\star}), \\[4pt]
\hat{d}_i^{\,2} &= \frac{\tilde{\beta}_i}{\tilde{\alpha}_i}, \qquad 
\hat{f}_1 = 0, \qquad v_1 = 0.
\end{aligned}
\end{equation}

When predicting future observations based on partially observed spatio-temporal data, we utilize the posterior conditional distribution to sequentially forecast future states. Suppose we have observed all data up to a specific cutoff time $t_0$ and wish to predict the remaining field for $t > t_0$. Because the time-ordering scheme naturally sorts the data chronologically, the indexing is aligned with predicting into the future and hence, the above posterior distribution can be applied sequentially at increasing $i$. Let $N_0$ denote the total number of observed responses up to time $t_0$ (i.e., $N_0 = |\{i : t_i \leq t_0\}|$). Given these observed values $\mathbf{y}_{1:N_0}^\ast$, the predictive distribution is:
\begin{equation}
p(\mathbf{y}_{(N_0 + 1):N}^\ast \mid \mathbf{y}_{1:N_0}^\ast, \hat{\boldsymbol{\theta}})
= 
\prod_{i = N_0 + 1}^{N}
t_{2\tilde{\alpha}_i}
\!\left(
y_i^\ast \,\middle|\, 
\hat{f}_i(\mathbf{y}_{1:i-1}^\ast),\,
\hat{d}_i^2 \big(v_i(\mathbf{y}_{1:i-1}^\ast) + 1 \big)
\right).
\end{equation}

Algorithm \ref{alg:full} concludes this section by providing the full procedure of our method.

\begin{algorithm}[H] 
\caption{Spatio-temporal transport map inference}
\label{alg:full}
\begin{enumerate}
    \item \textbf{Data preprocessing:} Standardize $\bX= [\bx_1, \ldots, \bx_N]^\top$ to zero mean and unit variance for each covariate.

    \item \textbf{Initial Range Estimation: } Estimate the spatial and temporal ranges by averaging the outputs of Algorithm~\ref{alg:length_scales}:
$
\lambda_s = \frac{1}{w} \sum_{k=1}^{w} \lambda_{s,k}, \;
\lambda_t = \frac{1}{w} \sum_{k=1}^{w} \lambda_{t,k}.
$

    \item \textbf{Ordering and Neighborhood Construction: } Find order \textit{ord} of $\tilde{\bX} = [\tilde{\bx}_1, \ldots, \tilde{\bx}_N]$ using either (1) maximin ordering or (2) time ordering (see Section \ref{sec:ord_and_cond}). $\bY \leftarrow \bY[ord]$. Calculate \rev{$l_i = \min_{j = 1, \ldots, i -1}\|\tilde{\bx}_i - \tilde{\bx}_j\|$ and $c_m(i)$}, $\forall i = 1, \ldots, N$.

    \item \textbf{Model Training:} Compute $\hat\bftheta = \arg\max_{\boldsymbol\theta} \log p(\mathbf{Y \mid \boldsymbol\theta })$ via gradient-based algorithm.

    \item \textbf{Sampling.}  
    Use the trained transport map to generate synthetic realizations at the target spatio-temporal locations, either (i) conditionally on partially observed data or (ii) unconditionally.

\end{enumerate}
\end{algorithm}

\section{Applications to climate data} \label{sec:application}

Climate models are computer programs based on complex mathematical representations of the Earth's climate system. They use systems of differential equations to simulate intricate interactions among the atmosphere, oceans, land surface, and ice. While these models are crucial for understanding climate dynamics and projecting future scenarios, developing and running them requires immense computational resources. For example, \citet{Dennis2012CESMPerformance} conducted performance studies on the Community Earth System Model (CESM), demonstrating that high-fidelity simulations require tens of thousands of processor cores and significant wall-clock time. Consequently, our proposed \rev{stochastic} modeling framework can be used to accurately infer the joint distribution of available climate ensembles. Once trained, the model acts as a highly efficient statistical surrogate, capable of generating new synthetic ensembles at a fraction of the computational cost of the original physics-based simulators.

To demonstrate the capability and scalability of our approach, we apply it to two non-Gaussian climate datasets obtained from the CESM Large Ensemble Project \citep{Kay2015, Hurrell2013CESM}:
\begin{enumerate}
    \item \textbf{Regional Precipitation:} The log-transformed total precipitation rate (in m/s) over Central America. The data are observed on a regular longitude--latitude grid of size $N_s = 74 \times 37 = 2,738$. 
    \item \textbf{Global Surface Temperature:} Global surface temperature data observed on a massive longitude--latitude grid of size $N_s = 288 \times 192 = 55,296$. 
\end{enumerate}
For both datasets, the temporal domain spans a 30-day period from July 1 to July 30 (\rev{$t = 1, \ldots, 30$, so that $N_t = 30$}). A total of $n=98$ independent realizations (ensembles) are available for each dataset, corresponding to 98 consecutive years beginning in model year 402 of a pre-industrial control simulation. \rev{Fewer than $1\%$ of the recorded precipitation values are exactly zero, which likely reflects the drizzle bias common to climate models. We floor these values at $10^{-10}$ m/s before taking logarithms, so the point mass at zero is negligible and we treat precipitation as a continuous field.}

There are two primary goals in this study: (1) to learn the full joint distribution of the spatio-temporal field $\by$ for unconditional sample generation, and (2) to learn the conditional distribution of future states given past observations (i.e., forecasting $\by_{t_{\star} + 1}, \ldots, \by_{N_t}$ given $\by_{1}, \ldots, \by_{t_{\star}}$). To achieve these goals, we evaluate the performance of our ARGP transport map under both the maximin space-time ordering and the time ordering. We benchmark our method against a standard parametric GP employing the Vecchia approximation. 

Model performance is evaluated using the average log-score on a hold-out test set of 10 ensemble members. The log-score, defined as the negative log-predictive density evaluated at the true test values, is a strictly proper scoring rule. It rigorously evaluates both the accuracy of the point predictions and the calibration of the predictive uncertainty. Up to an additive constant, minimizing the log-score is equivalent to minimizing the Kullback-Leibler (KL) divergence between the true data-generating distribution and the modeled distribution. To assess sample efficiency, we also vary the number of training ensembles from 10 to 80 in increments of 10.

\subsection{Data processing}

To ensure numerical stability and rotational invariance across the globe, we first convert the geographic longitude--latitude coordinates into 3D Cartesian coordinates ($x, y, z$) \rev{on a sphere of radius 1}. This transformation elegantly circumvents artificial boundary discontinuities at the dateline and poles. We subsequently standardize both the spatial and temporal covariates \rev{($x, y, z, t$)} to have a mean of zero and a standard deviation of one. Finally, the climate responses (log-precipitation and temperature) are standard-scaled at each specific spatio-temporal location across the training ensembles, using the location-wise mean $\hat\mu_i$ and standard deviation $\hat\sigma_i$. \rev{For log-precipitation, the inverse-transform $y \mapsto \exp(\hat\sigma_i y + \hat\mu_i)$ is a monotone bijection onto $(0,\infty)$, so every draw from the predictive distribution corresponds to a strictly positive precipitation rate.}

Following the procedure outlined in Section \ref{sec: estimation}, we estimate the spatial and temporal length-scale parameters prior to fitting the transport map. To maintain computational efficiency, we fit a parametric GP using a randomly sampled subset of 5,000 data points across 5 ensembles. We assume an \rev{Automatic Relevance Determination (ARD) \citep{WilliamsARD1995}} Mat\'ern covariance kernel with a smoothness parameter of $\nu = 1.5$, \rev{allowing separate length scales for the spatial and temporal covariates, with the three spatial coordinates constrained to share a common scale}. In our experiments, increasing the subset size beyond 5,000 did not yield significant differences in the estimated scale parameters. \rev{The standard deviations of the estimated length scales across the repeated subsets are of order $10^{-5}$. The results are also insensitive to the assumed smoothness: re-estimating the length scales with $\nu = 0.5$ and propagating the resulting scaling through the ordering and the conditioning sets leaves the log-scores essentially unchanged at every training ensemble size.} \rev{Constructing the ordering and the conditioning sets takes less than a minute even for the largest dataset considered here, and accounts for a small fraction of the total fitting time.}

\subsection{Application to regional precipitation data} \label{sec:precip_app}

\begin{figure}[ht]
\centering
    \begin{subfigure}{1.0\textwidth}
    \centering
    \includegraphics[clip, width =.98\linewidth]{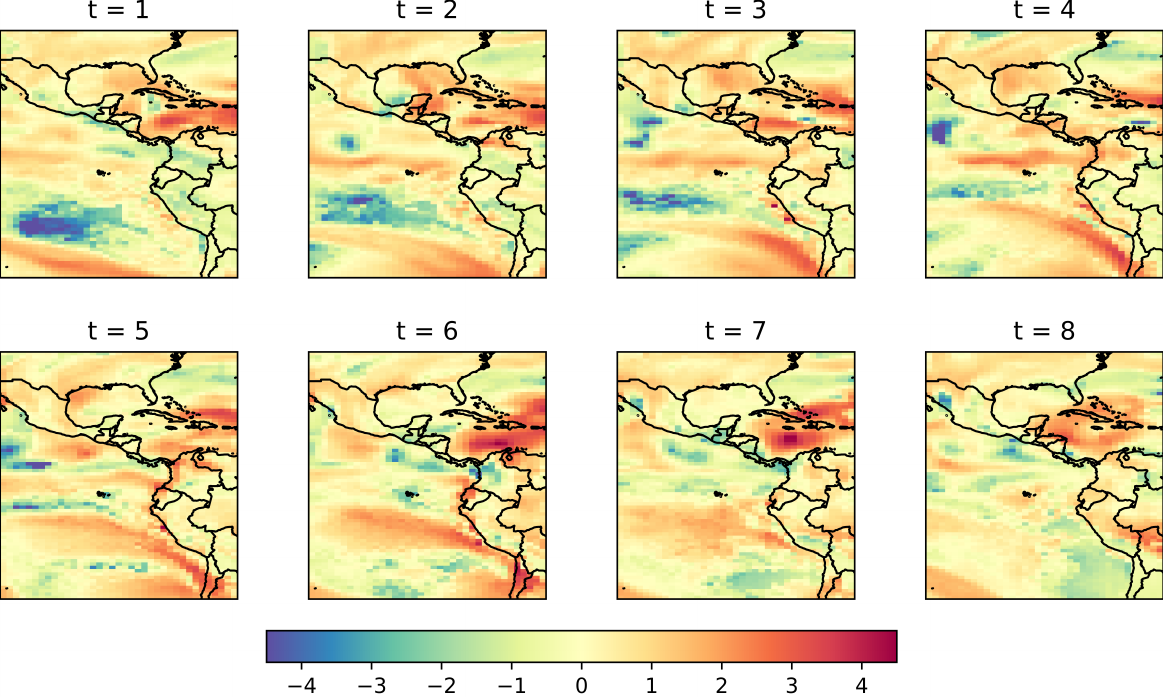}
    \end{subfigure}
    \caption{Visualization of one ensemble member from the Central America regional precipitation dataset. Each panel represents the spatial field at a single time point (day).}
    \label{fig:real_precip}
\end{figure}

Our first dataset focuses on daily precipitation during July over Central America, corresponding to a latitude range of $[29.69^\circ S, 39.11^\circ N]$ and a longitude range of $[250^\circ, 295^\circ]$. Given the $2,738$ spatial locations and $30$ time points, each independent ensemble contains $N = 2,738 \times 30 = 82,140$ data points. The response variable $\mathbf{Y}$ represents the pre-processed daily precipitation, while the input covariates consist of the 3D Cartesian spatial coordinates \rev{on a sphere of radius 1} and the day of the month. Figure \ref{fig:real_precip} visualizes a sequence of time frames from one of the true ensemble members.

Under our pre-processing described above (with all coordinates scaled to unit variance), the estimated length scales for the spatial and temporal inputs were found to be 0.427 and 0.320, respectively. These scales were used to compute the scaled pairwise distances, allowing us to perform the spatio-temporal maximin ordering and construct the local conditioning sets. The size of the conditioning set $m$ optimized by our model was approximately 22, varying slightly (between 20 and 24) depending on the exact training size and chosen ordering method. 

\begin{figure}[ht]
\centering
    \begin{subfigure}{.75\textwidth}
    \centering
    \includegraphics[clip, width =.98\linewidth]{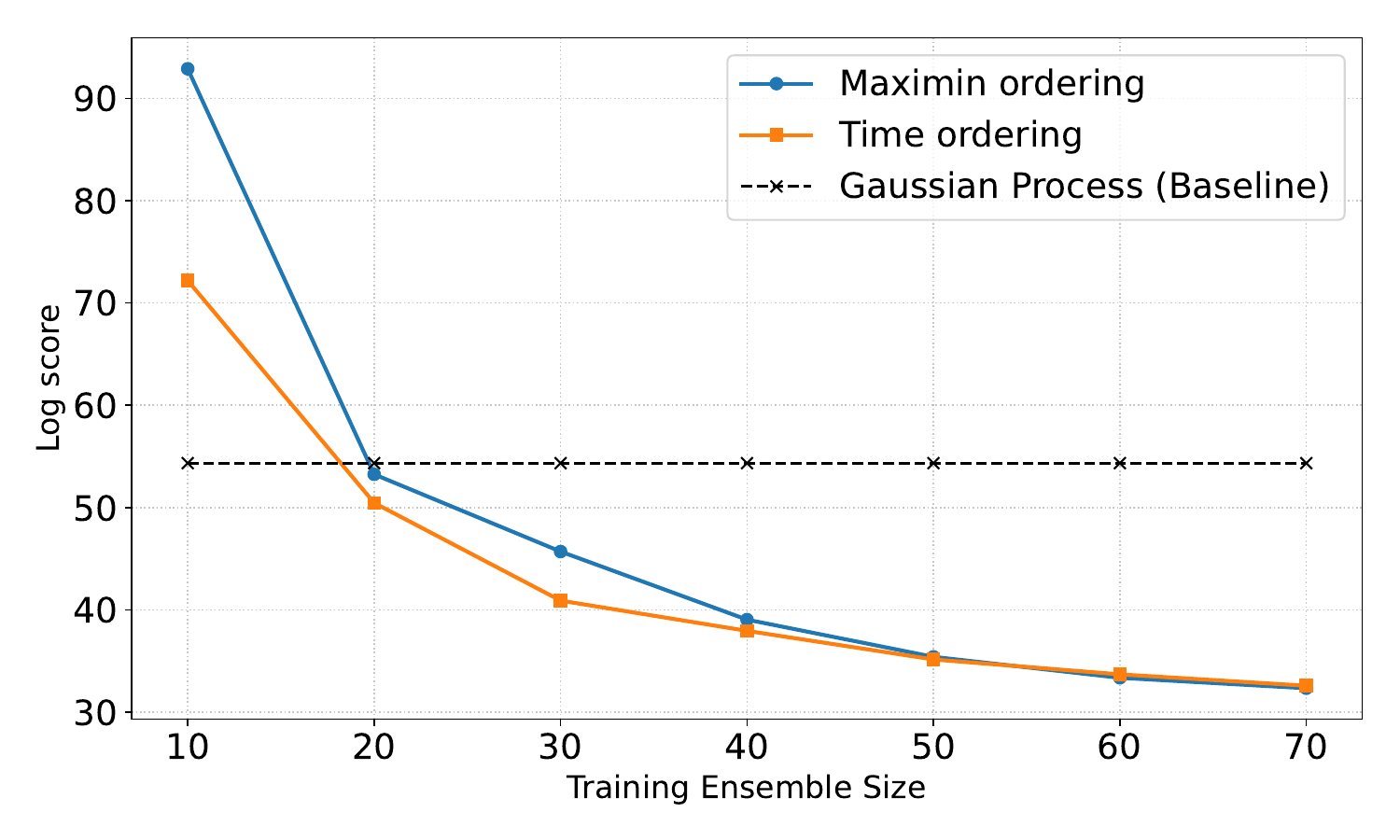}
    \end{subfigure}
    \caption{Log-score \rev{(in thousands)} comparison between our ARGP framework \rev{(maximin and time ordering)} and a parametric GP model as a function of training ensemble size. The parametric GP does not improve substantially with increasing sample size. Our ARGP achieves lower (better) log-scores when trained on twenty samples or more.}
    \label{fig:precip_logscores}
\end{figure}

Figure \ref{fig:precip_logscores} presents the log-score comparison between our trained transport map and a parametric Vecchia GP model on the 10 hold-out test samples. The baseline parametric GP was unable to achieve log-scores below $50{,}000$. In stark contrast, our ARGP approach achieved significantly improved log-scores of approximately $32{,}600$ when utilizing the full training set. Furthermore, as the training size increases, the predictive performance of our model consistently improves, surpassing the parametric GP benchmark once the training size reaches 20 ensembles. This indicates that our non-Gaussian transport map captures the complex, underlying spatio-temporal distribution much more accurately than parametric linear Gaussian assumptions.

\subsection{Application to global surface temperature data}

To test the extreme scalability of our framework, we analyzed the global surface temperature dataset. With 98 ensembles measured over a 288-by-192 spatial grid across 30 days, each individual ensemble member comprises $N = 1,658,880$ spatio-temporal data points. 

Under our pre-processing described above (with all coordinates scaled to unit variance), the estimated length-scale parameters for the four covariates (3D spatial and 1D temporal) were (0.67,0.67,0.67,0.87). Note that the three spatial Cartesian covariates were constrained to share a common length scale to ensure isotropic spatial distances. We then scaled the coordinates, established the orderings, and successfully fit the transport-map model using training ensemble sizes ranging from 10 to 80. \rev{The optimized conditioning set size $m$ was determined to be approximately 16, smaller than for the regional precipitation data. This is consistent with the stronger dependence of the surface-temperature field, for which fewer neighbors are needed to capture the conditional structure.}

\begin{figure}[ht]
\centering
    \begin{subfigure}{1\textwidth}
    \centering
    \includegraphics[width=0.95\textwidth]{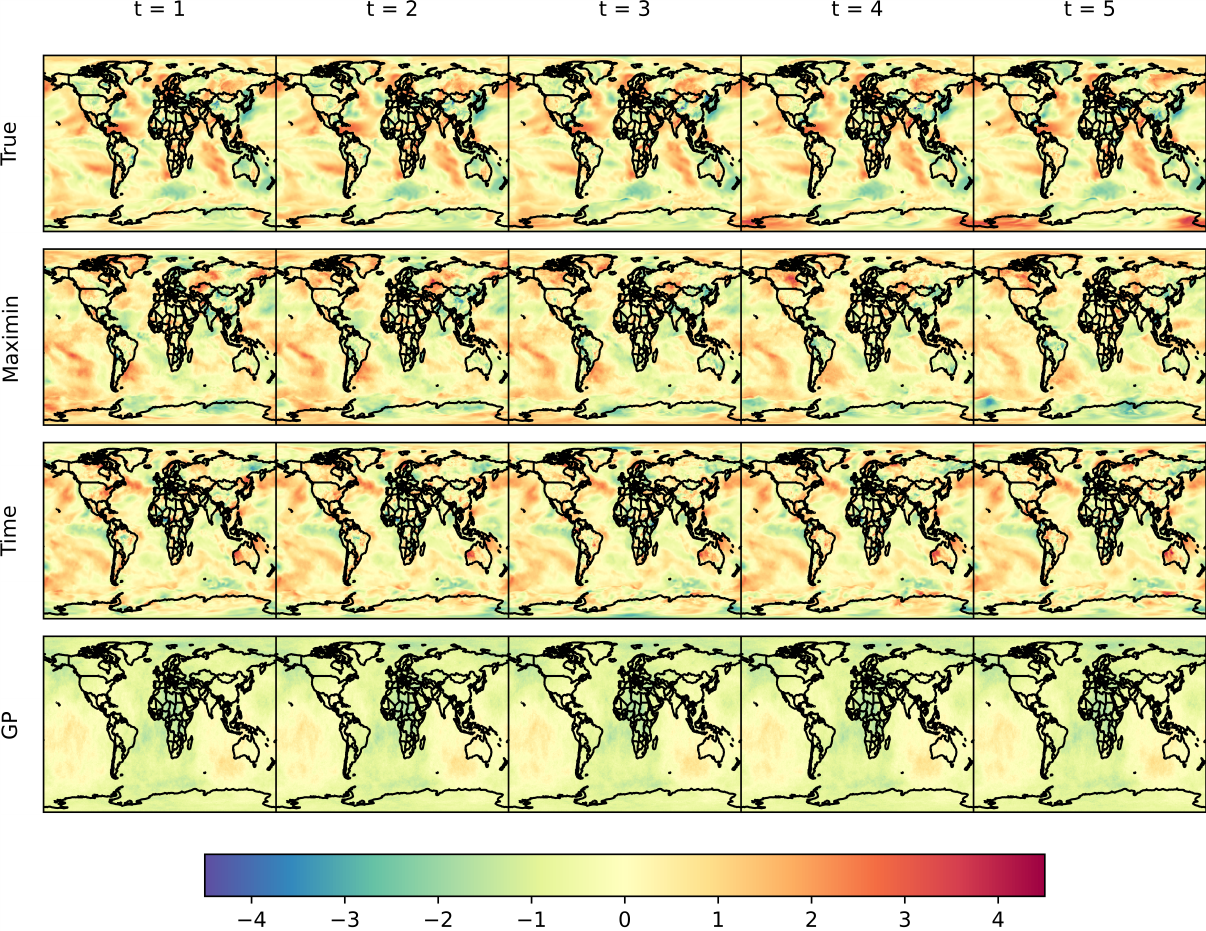}
    \end{subfigure}
    \caption{Unconditional sample generation. Top row: true unobserved (test) global surface temperature fields. Middle row: synthetic samples produced by our ARGP method (using global maximin ordering). Bottom row: synthetic samples produced by the baseline parametric GP. \rev{Columns show spatial snapshots of the temperature field on five consecutive days, $t = 1, \ldots, 5$, as standardized anomalies.}}
    \label{fig:ls_precip_global_maxmin}
\end{figure}

For baseline comparison, we trained a parametric GP model using the \texttt{matern\_spacetime} kernel from the \texttt{GpGp} R package. \rev{The baseline was given the same ordering and the same conditioning sets as our model, so that the two methods differ only in how the conditional distributions are specified; its covariance hyperparameters were then estimated by maximum likelihood within \texttt{GpGp}.} Figure \ref{fig:ls_precip_global_maxmin} provides a visual comparison of unconditional samples generated by our model (using maximin ordering) and the parametric GP, evaluated against the true observations. The fields generated by our ARGP model exhibit spatial and temporal textures that are visibly more consistent with the true physics-based simulator.

\begin{figure}[ht]
\centering
    \begin{subfigure}{1\textwidth}
    \centering
    \includegraphics[clip, width =.98\linewidth]{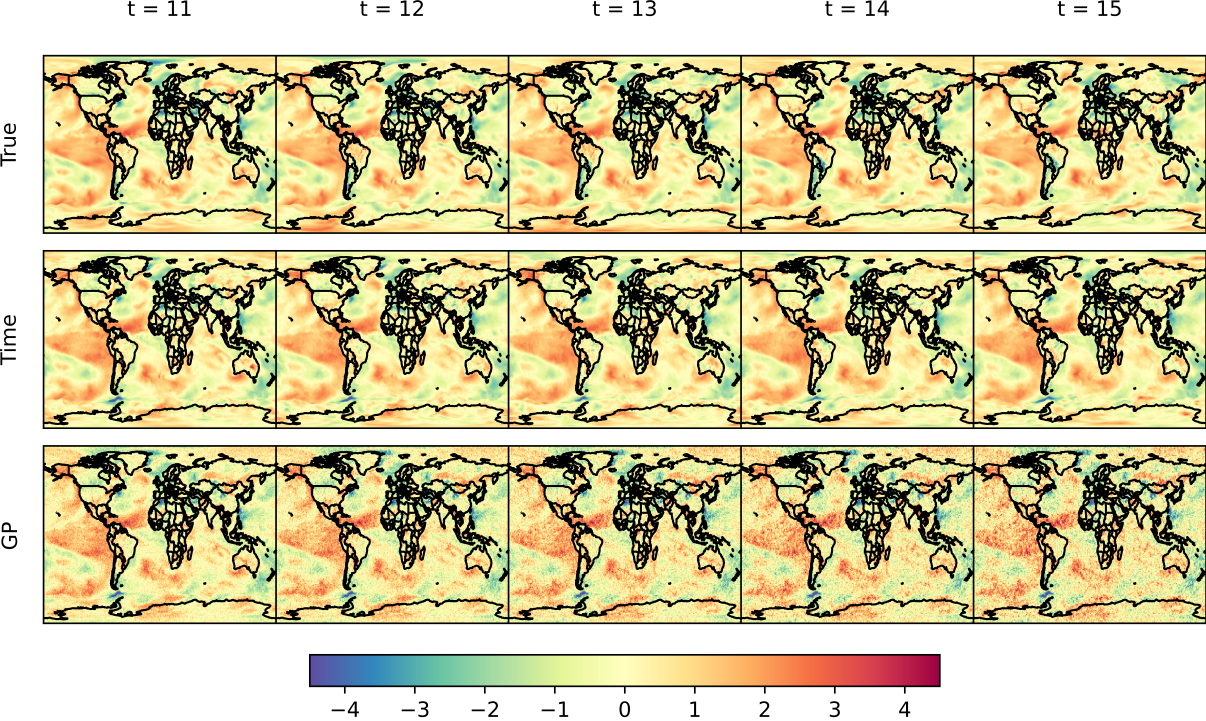}
    \end{subfigure}
    \caption{Conditional temporal forecasts given the first 10 time frames. Top row: true unobserved test data trajectory. Middle row: conditional forecast produced by our ARGP method (using time ordering). Bottom row: conditional forecast produced by the parametric GP. \rev{Columns show predicted spatial snapshots of the temperature field on five consecutive forecast days, $t = 11, \ldots, 15$, as standardized anomalies.}}
    \label{fig:global_condsamp_compar}
\end{figure}

\begin{figure}[ht]
\centering
    \begin{subfigure}{.7\textwidth}
    \centering
    \includegraphics[clip, width =.98\linewidth]{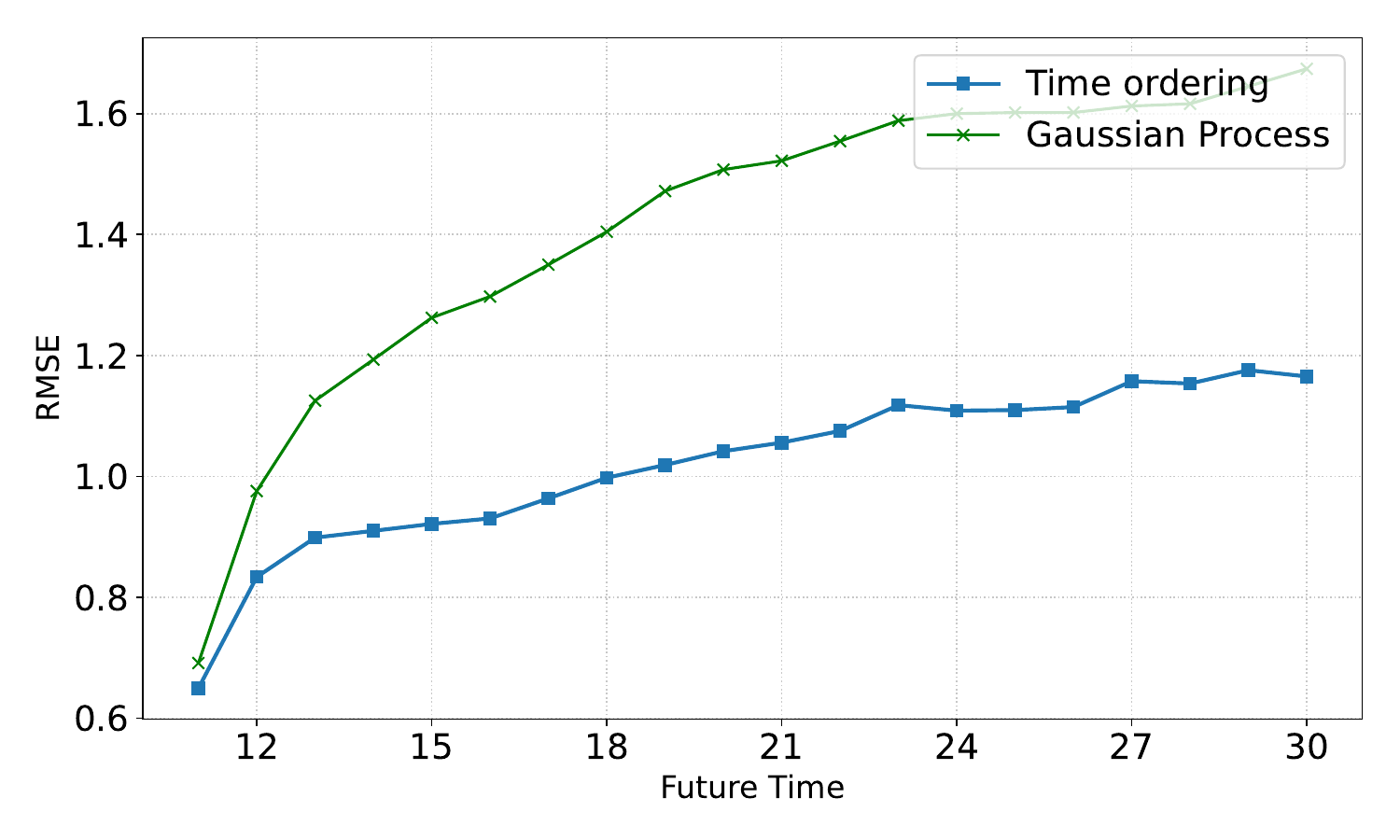}
    \end{subfigure}
    \caption{\rev{Spatially averaged root-mean-square error (RMSE) at each lead time for our ARGP method under time ordering and for the parametric GP, computed over 100 samples conditioned on the first 10 time frames.}}
    \label{fig:rmse_leadtime}
\end{figure}

Finally, we utilized the time-ordering scheme to evaluate conditional forecasting. Based on the trained model, we provided the first 10 time frames of a test ensemble as the conditioning set $\mathbf{y}_{1:N_0}^\ast$ and drew samples to "forecast" the remaining time frames. Figure \ref{fig:global_condsamp_compar} illustrates that, given identical historical data, our method produces future spatial fields that align much more closely with the actual ground-truth trajectory compared to the parametric GP forecast. \rev{Figure \ref{fig:rmse_leadtime} compares the spatially averaged root-mean-square error (RMSE) at each lead time, computed over 100 conditional samples. Our method achieves lower RMSE than the parametric GP across the forecast horizon, and the gap does not close as the forecast extends.}

\section{Discussion and future work} \label{sec:discussion}
In this paper, we introduced a highly scalable, \rev{stochastic} modeling framework for non-Gaussian spatio-temporal fields using autoregressive Gaussian processes (ARGPs). By representing the joint distribution as a product of univariate conditional distributions, our approach circumvents the limitations of traditional linear Gaussian assumptions without requiring the massive training datasets typically demanded by deep-learning models. Because the nonparametric Bayesian procedure rigorously quantifies uncertainty and employs data-dependent sparsity, it is highly sample-efficient and exceptionally scalable. We successfully demonstrated our method on a complex spatio-temporal climate-model field in over 1.6 million dimensions, utilizing only a small ensemble of training samples.

A key methodological innovation of our work is the introduction of a scaled spatio-temporal coordinate metric, which automatically balances spatial and temporal dependencies to determine optimal conditioning sets. Furthermore, we showed that the ARGP framework can be tailored to specific predictive tasks through the choice of ordering schemes. The global maximin space-time ordering proved highly effective for learning the full joint distribution and generating new climate fields. Conversely, our novel time-ordering scheme naturally partitions historical and future data, enabling accurate, sequential conditional forecasting of future spatial trajectories. \rev{In the regional precipitation application, our model attained substantially better predictive log-scores than a standard parametric Vecchia-approximated GP, and in the global temperature application its samples and forecasts were visibly closer to the simulator output.}

Looking forward, there are several promising avenues for extending this framework. We recently introduced a multi-resolution and downscaling extension of the ARGP framework for purely spatial fields \citep{Calle-Saldarriaga2025}. A natural next step is to extend this multi-resolution capability to the spatio-temporal domain, allowing the model to capture different dynamical structures at different scales. In climate modeling, physical processes often exhibit varying dependence structures depending on the resolution (e.g., broad atmospheric wave dynamics versus localized, rapidly evolving convective precipitation). Adapting the spatio-temporal ARGP to fuse low-fidelity, coarse-resolution global models with high-fidelity, fine-resolution regional models could further enhance predictive accuracy and computational efficiency. \rev{Computationally, the factorization of the likelihood into $N$ independent terms of identical form, each requiring an $n \times n$ Cholesky factorization, makes our method well suited to GPU acceleration \citep{james2024,gyger2026scalablenonseparablespatiotemporalgaussian}. The ordering and conditioning-set construction are inherently sequential, as is sampling within a single realization, but both amount to a minor cost compared to model estimation. Furthermore, sampling parallelizes across replicates.} \rev{A future direction is to recenter the prior on a parametric spatio-temporal target, extending to space-time the parametric shrinkage that \citet{chakraborty2025} develop for purely spatial transport maps, and which goes back to the targeted covariance estimation of \citet{kidd2022}. We expect this to be helpful for small training ensembles.}

Additionally, future work could explore integrating physical constraints directly into the GP prior mean functions to ensure that generated samples rigorously adhere to conservation laws (e.g., mass and energy balance) over time. Finally, while our application focused on climate-model emulation, the extreme scalability and flexibility of the proposed spatio-temporal ARGP make it a highly promising tool for other high-dimensional domains, such as epidemiology, oceanography, and fluid dynamics.

\ifcopernicus
\else

\section*{Code and Data Availability}
\rev{Code implementing the proposed method, together with a demonstration using
the data from the application, is archived on Zenodo
\citep{leicramer_2026_stargp} and available at
\url{https://github.com/katzfuss-group/space_time_demo} under the MIT
License. The global surface temperature data from the CESM Large Ensemble
Project \citep{Kay2015, Hurrell2013CESM} used in the application are archived on
Zenodo \citep{leicramer_2026_temperature} under a CC BY 4.0 License.}

\ifcopernicus
\section*{Author Contributions}
C.L.\ developed the methodology, implemented the model, performed the analyses, and wrote the manuscript.  
J.C.\ contributed to methodological discussions, provided technical guidance, and assisted in refining the analysis and manuscript.  
M.K.\ conceived the study, supervised the project, contributed to the methodological and conceptual development, and revised the manuscript.
\fi

\section*{Acknowledgements}
The authors were partially supported by NASA AIST-21/23 and NSF DMS--1953005/2433548. We thank Jonathan Hobbs for helpful comments. The authors used ChatGPT (OpenAI) and Claude Code (Anthropic) to assist with editing and proofreading. The authors reviewed and take full responsibility for the final content.

\fi

\ifcopernicus
\else
    \bibliographystyle{apalike}
\fi
\bibliography{mendeley,additionalrefs}

@article{chen2024precision,
  title={Precision and Cholesky Factor Estimation for Gaussian Processes},
  author={Chen, Jiaheng and Sanz-Alonso, Daniel},
  journal={arXiv preprint arXiv:2412.08820},
  year={2024}
}

@article{Dennis2012CESMPerformance,
  author    = {Dennis, John M. and Vertenstein, Mariana and Worley, Patrick H. and Mirin, Arthur A. and Craig, Anthony P. and Jones, Philip W. and Mickelson, Shawn A. and Jacob, Robert L.},
  title     = {Computational Performance of the Community Earth System Model},
  journal   = {International Journal of High Performance Computing Applications},
  volume    = {26},
  number    = {1},
  pages     = {5--16},
  year      = {2012},
  doi       = {10.1177/1094342011428143},
  publisher = {SAGE Publications}
}

@article{Hurrell2013CESM,
  title={The Community Earth System Model: A framework for collaborative research},
  author={Hurrell, James W. and Holland, Marika M. and Gent, Peter R. and Ghan, Steven and Kay, Jennifer E. and Kushner, Paul J. and Lamarque, Jean-François and Large, William G. and Lawrence, David and Lindsay, Keith and others},
  journal={Bulletin of the American Meteorological Society},
  volume={94},
  number={9},
  pages={1339--1360},
  year={2013},
  doi={10.1175/BAMS-D-12-00121.1}
}

@article{Calle-Saldarriaga2025,
    title = {{Generative multi-fidelity modeling and downscaling via spatial autoregressive transport maps}},
    year = {2025},
    journal = {arXiv:2509.22474},
    author = {Calle-Saldarriaga, Alejandro and Wiemann, Paul FV and Katzfuss, Matthias}
}

@inproceedings{WilliamsARD1995,
 author = {Williams, Christopher and Rasmussen, Carl},
 booktitle = {Advances in Neural Information Processing Systems},
 editor = {D. Touretzky and M.C. Mozer and M. Hasselmo},
 pages = {},
 publisher = {MIT Press},
 title = {Gaussian Processes for Regression},
 url = {https://proceedings.neurips.cc/paper_files/paper/1995/file/7cce53cf90577442771720a370c3c723-Paper.pdf},
 volume = {8},
 year = {1995}
}

@article{Katzfuss02042024,
author = {Matthias Katzfuss and Florian Schäfer},
title = {{Scalable Bayesian Transport Maps for High-Dimensional Non-Gaussian Spatial Fields}},
journal = {Journal of the American Statistical Association},
volume = {119},
number = {546},
pages = {1409--1423},
year = {2024},
publisher = {Taylor \& Francis},
doi = {10.1080/01621459.2023.2197158},
URL = {
https://doi.org/10.1080/01621459.2023.2197158
},
eprint = {
https://doi.org/10.1080/01621459.2023.2197158
}
}

@misc{brachem2026dataefficientgenerativemodelingnongaussian,
title={{Data-Efficient Generative Modeling of Non-Gaussian Global Climate Fields via Scalable Composite Transformations}},
author={Johannes Brachem and Paul F. V. Wiemann and Matthias Katzfuss},
year={2026},
eprint={2602.23311},
archivePrefix={arXiv},
primaryClass={stat.ME},
url={https://arxiv.org/abs/2602.23311},
}

@misc{gyger2026scalablenonseparablespatiotemporalgaussian,
title={Scalable non-separable spatio-temporal {G}aussian process models for large-scale short-term weather prediction},
author={Tim Gyger and Reinhard Furrer and Fabio Sigrist},
year={2026},
eprint={2602.03609},
archivePrefix={arXiv},
primaryClass={stat.AP},
url={https://arxiv.org/abs/2602.03609},
}

@article{wiemann2023,
author = {Wiemann, Paul F. V. and Katzfuss, Matthias},
date = {2023/12/01},
doi = {10.1007/s13253-023-00580-z},
id = {Wiemann2023},
isbn = {1537-2693},
journal = {Journal of Agricultural, Biological and Environmental Statistics},
number = {4},
pages = {597--617},
title = {{Bayesian Nonparametric Generative Modeling of Large Multivariate Non-Gaussian Spatial Fields}},
url = {https://doi.org/10.1007/s13253-023-00580-z},
volume = {28},
year = {2023}}

@article{chakraborty2025,
author = {Chakraborty, Anirban and Katzfuss, Matthias},
date = {2025/03/22},
doi = {10.1007/s13253-025-00687-5},
id = {Chakraborty2025},
isbn = {1537-2693},
journal = {Journal of Agricultural, Biological and Environmental Statistics},
title = {{Learning Non-Gaussian Spatial Distributions Via Bayesian Transport Maps with Parametric Shrinkage}},
url = {https://doi.org/10.1007/s13253-025-00687-5},
year = {2025}}

@article{james2024,
author = {James, Zachary and Guinness, Joseph},
date = {2024/10/28},
doi = {10.1007/s11222-024-10510-9},
id = {James2024},
isbn = {1573-1375},
journal = {Statistics and Computing},
number = {6},
pages = {207},
title = {{Implementation and analysis of GPU algorithms for Vecchia Approximation}},
url = {https://doi.org/10.1007/s11222-024-10510-9},
volume = {34},
year = {2024}}

@misc{idir2025improvingspatiotemporalgaussianprocess,
title={{Improving Spatio-temporal Gaussian Process Modeling with Vecchia Approximation: A Low-Cost Sensor-Driven Approach to Urban Environmental Monitoring}},
author={Yacine Mohamed Idir and Olivier Orfila and Patrice Chatellier and Vincent Judalet},
year={2025},
eprint={2511.22500},
archivePrefix={arXiv},
primaryClass={stat.ME},
url={https://arxiv.org/abs/2511.22500},
}

@article{kidd2022,
author = {Brian Kidd and Matthias Katzfuss},
title = {{Bayesian Nonstationary and Nonparametric Covariance Estimation for Large Spatial Data (with Discussion)}},
volume = {17},
journal = {Bayesian Analysis},
number = {1},
publisher = {International Society for Bayesian Analysis},
pages = {291 -- 351},
year = {2022},
doi = {10.1214/21-BA1273},
URL = {https://doi.org/10.1214/21-BA1273}
}

@misc{leicramer_2026_stargp,
  author    = {Lei-Cramer, Carrie and Cao, Jian and Katzfuss, Matthias},
  title     = {Scalable spatio-temporal autoregressive Gaussian processes:
               code},
  year      = {2026},
  month     = sep,
  publisher = {Zenodo},
  version   = {v1.0.0},
  doi       = {10.5281/zenodo.22882806},
  url       = {https://doi.org/10.5281/zenodo.22882806}
}

@dataset{leicramer_2026_temperature,
  author    = {Lei-Cramer, Carrie and Cao, Jian and Katzfuss, Matthias},
  title     = {Global surface temperature fields from the {CESM} Large
               Ensemble for non-{G}aussian spatio-temporal modeling},
  year      = {2026},
  month     = sep,
  publisher = {Zenodo},
  version   = {v1},
  doi       = {10.5281/zenodo.22881514},
  url       = {https://doi.org/10.5281/zenodo.22881514}
}

@article{Katzfuss2017a,
    title = {{A general framework for Vecchia approximations of Gaussian processes}},
    year = {2021},
    journal = {Statistical Science},
    author = {Katzfuss, Matthias and Guinness, Joseph},
    number = {1},
    pages = {124--141},
    volume = {36},
    url = {http://arxiv.org/abs/1708.06302},
    doi = {10.1214/19-STS755},
    arxivId = {1708.06302}
}

@article{Kovachki2020,
    title = {{Conditional sampling with monotone GANs}},
    year = {2020},
    journal = {arXiv:2006.06755},
    author = {Kovachki, Nikola B. and Hosseini, Bamdad and Baptista, Ricardo and Marzouk, Youssef M.},
    issn = {23318422},
    arxivId = {2006.06755}
}

@book{Goodfellow2016,
    title = {{Deep Learning}},
    year = {2016},
    author = {Goodfellow, Ian and Bengio, Yoshua and Courville, Aaron},
    publisher = {MIT Press}
}

@article{Hestness2017,
    title = {{Deep learning scaling is predictable, empirically}},
    year = {2017},
    journal = {arXiv:1712.00409},
    author = {Hestness, Joel and Narang, Sharan and Ardalani, Newsha and Diamos, Gregory and Jun, Heewoo and Kianinejad, Hassan and Patwary, Md. Mostofa Ali and Yang, Yang and Zhou, Yanqi},
    url = {http://arxiv.org/abs/1712.00409},
    arxivId = {1712.00409}
}

@article{Vecchia1988,
    title = {{Estimation and model identification for continuous spatial processes}},
    year = {1988},
    journal = {Journal of the Royal Statistical Society, Series B},
    author = {Vecchia, AV},
    number = {2},
    pages = {297--312},
    volume = {50},
    url = {http://www.jstor.org/stable/10.2307/2345768}
}

@article{Choi2013,
    title = {{Nonparametric estimation of spatial and space-time covariance function}},
    year = {2013},
    journal = {Journal of Agricultural, Biological, and Environmental Statistics},
    author = {Choi, InKyung K. and Li, Bo and Wang, Xiao},
    number = {4},
    pages = {611--630},
    volume = {18},
    doi = {10.1007/s13253-013-0152-z},
    issn = {10857117}
}

@article{Huang2011,
    title = {{Nonparametric estimation of the variogram and its spectrum}},
    year = {2011},
    journal = {Biometrika},
    author = {Huang, Chunfeng and Hsing, Tailen and Cressie, Noel},
    number = {4},
    month = {11},
    pages = {775--789},
    volume = {98},
    url = {http://biomet.oxfordjournals.org/cgi/doi/10.1093/biomet/asr056},
    doi = {10.1093/biomet/asr056},
    issn = {00063444}
}

@article{Guinness2016a,
    title = {{Permutation and Grouping Methods for Sharpening Gaussian Process Approximations}},
    year = {2018},
    journal = {Technometrics},
    author = {Guinness, Joseph},
    number = {4},
    month = {10},
    pages = {415--429},
    volume = {60},
    url = {http://arxiv.org/abs/1609.05372 https://www.tandfonline.com/doi/full/10.1080/00401706.2018.1437476},
    doi = {10.1080/00401706.2018.1437476},
    issn = {0040-1706},
    arxivId = {1609.05372}
}

@incollection{Marzouk2016,
    title = {{Sampling via measure transport: An introduction}},
    year = {2016},
    booktitle = {Handbook of Uncertainty Quantification},
    author = {Marzouk, Youssef M. and Moselhy, Tarek and Parno, Matthew and Spantini, Alessio},
    editor = {Ghanem, R and Higdon, Dave and Owhadi, Houman},
    publisher = {Springer},
    isbn = {9783319123851},
    doi = {10.1007/978-3-319-12385-1}
}

@article{Schafer2020,
    title = {{Sparse Cholesky factorization by Kullback-Leibler minimization}},
    year = {2021},
    journal = {SIAM Journal on Scientific Computing},
    author = {Sch{\"{a}}fer, Florian and Katzfuss, Matthias and Owhadi, Houman},
    number = {3},
    pages = {A2019-A2046},
    volume = {43},
    doi = {10.1137/20M1336254},
    arxivId = {2004.14455}
}

@book{Cressie1993,
    title = {{Statistics for Spatial Data, revised edition}},
    year = {1993},
    author = {Cressie, Noel},
    edition = {},
    publisher = {John Wiley {\&} Sons},
    url = {http://books.google.com/books?id=4SdRAAAAMAAJ&pgis=1},
    address = {New York, NY}
}

@article{Kay2015,
    title = {{The Community Earth System Model (CESM) Large Ensemble Project: A community resource for studying climate change in the presence of internal climate variability}},
    year = {2015},
    journal = {Bulletin of the American Meteorological Society},
    author = {Kay, J. E. and Deser, C. and Phillips, A. and Mai, A. and Hannay, C. and Strand, G. and Arblaster, J. M. and Bates, S. C. and Danabasoglu, G. and Edwards, J. and Holland, M. and Kushner, P. and Lamarque, J. F. and Lawrence, D. and Lindsay, K. and Middleton, A. and Munoz, E. and Neale, R. and Oleson, K. and Polvani, L. and Vertenstein, M.},
    number = {8},
    pages = {1333--1349},
    volume = {96},
    doi = {10.1175/BAMS-D-13-00255.1},
    issn = {00030007}
}

@inproceedings{Arjovsky2017,
    title = {{Towards principled methods for training generative adversarial networks}},
    year = {2017},
    booktitle = {International Conference on Learning Representations},
    author = {Arjovsky, Martin and Bottou, Léon},
    arxivId = {1701.04862}
}

@article{Stein2011,
    title = {{When does the screening effect hold?}},
    year = {2011},
    journal = {Annals of Statistics},
    author = {Stein, Michael L.},
    number = {6},
    month = {12},
    pages = {2795--2819},
    volume = {39},
    url = {http://projecteuclid.org/euclid.aos/1327413769},
    doi = {10.1214/11-AOS909},
    issn = {0090-5364}
}

\end{document}